\documentclass[aps,prl,twocolumn,superscriptaddress]{revtex4-2}
\usepackage{graphicx}
\usepackage{amsfonts}
\usepackage{amsmath}
\usepackage{amssymb}
\usepackage{bm}
\usepackage{hyperref}
\usepackage{slashed}
\usepackage{color}
\usepackage{ulem}

\begin{document}
\title{Unique quantum impurity states driven by a vortex in topological superconductors}

\author {Wei Su}
\affiliation{Department of Physics and Astronomy, Shanghai Jiao Tong University, Shanghai 200240, China}
\affiliation{College of Physics and Electronic Engineering, Center for Computational Sciences, Sichuan Normal University, Chengdu 610068, China}
\affiliation{Beijing Computational Science Research Center, Beijing 100084, China}

\author{Rui Wang}
\email{rwang89@nju.edu.cn}
\affiliation{National Laboratory of Solid State Microstructures and Department of Physics, Nanjing University, Nanjing 210093, China}
\affiliation{Collaborative Innovation Center for Advanced Microstructures, Nanjing 210093, China}

\author{Changfeng Chen}
\affiliation{Department of Physics and Astronomy, University of Nevada, Las Vegas, Nevada 89154, USA}

\author{Xiaoqun Wang}
\email{xiaoqunwang@sjtu.edu.cn}
\affiliation{Department of Physics and Astronomy, Shanghai Jiao Tong University, Shanghai 200240, China}
\affiliation{Collaborative Innovation Center for Advanced Microstructures, Nanjing 210093, China}
\affiliation{Tsung-Dao Lee Institute, Shanghai Jiao Tong University, Shanghai 200240, China}
\affiliation{Beijing Computational Science Research Center, Beijing 100084, China}

\date{\today}

\begin{abstract}
The interplay of magnetic impurity and vortex in a topological superconductor is of fundamental interest with major implications for implementing quantum computation. There are multiple degrees of freedom interacting with the impurity in the system, including the Majorana zero mode (MZM), the Caroli de Gennes Matricon (CdGM) states, and the electron bulk states that form Cooper pairs, which makes the impurity pinned vortex state elusive to date in topological superconductors. Here, we present an accurate solution of the problem, based on a generalized mapping scheme and the density-matrix renormalization group (DMRG) method.  We identify in-gap states that are distinct from the established Yu-Shiba-Rusinov (YSR) states. The newly found in-gap physics is driven by three prominent mechanisms: (i) the coupling of impurity and bulk states leading to competition between Kondo screening and Cooper pairing, resulting in a singlet-doublet quantum phase transition, (ii) the coupling of impurity and CdGM states introducing an effective Zeeman splitting to the doublet state, and (iii) the coupling of impurity and MZM turning the singlet-doublet transition into a crossover. These mechanisms cooperatively produce a unique spin-resolved local density of states. Despite the MZM, a robust nearly zero-energy peak is generated, with comparable height but opposite spin polarization as that of the MZM. These results signify novel emergent physics and offer insights to elucidate intriguing experimental phenomena.
\end{abstract}

\maketitle

\textcolor{blue}{\textit{Introduction.--}}
MZMs are self-conjugate particles obeying non-Abelian statistics, which in recent years have attracted great interest for their promise in implementing topological quantum computation \cite{RevModPhys.87.137,RevModPhys.80.1083}. Two-dimensional topological superconductors (TSCs) are widely believed as the suitable platform to host MZMs. Recent experimental progress suggests that such TSCs can be realized either in fabricated heterostructures between topological insulators and normal superconductors (SCs), or in certain materials that simultaneously hosts nontrivial band topology and intrinsic superconductivity, such as iron-based SCs, $\mathrm{FeTe}_{0.55}\mathrm{Se}_{0.45}$. In both cases, the Cooper pairs in the SC states proximate into the topological surface state with nontrivial spin textures, resulting in an effective $p+ip$ pairing symmetry that are conducive to hosting MZMs in the vortex regions. For clean TSCs in an external magnetic field, the zero-bias peaks (ZBPs) detected by STM were taken as the signature of MZMs \cite{PhysRevLett.126.127001, PhysRevLett.116.257003,PhysRevLett.117.047001, PhysRevX.8.041056, wang2018evidence,zhang2018observation}.

Impurities play important yet contrasting roles in influencing and probing pertinent physics in SCs. On the one hand, magnetic scatterings of the SC bath generate localized in-gap states, known as YSR states \cite{yu1965bound,shiba1968classical}, which offer insights into the nature of the SC states. Studies of both s-wave SC and TSCs without vortex reveal a quantum phase transition (QPT) between a singlet and doublet state, which is driven by competing Kondo scale and SC gap \cite{PhysRevLett.122.087001, PhysRevB.96.220507}. These states are experimentally accessible by STM measurements \cite{PhysRevLett.119.197002,cortes2021observation} and can be used to assess the underlying SC pairing symmetries. Moreover, MZMs can be produced and manipulated without an external magnetic field \cite{PhysRevX.9.011033, qin2020topological}. It has been proposed \cite{PhysRevX.9.011033} that a quantum anomalous vortex state can be induced by a classical magnetic impurity via the Elliot-Yafet spin-orbital coupling, and this scenario was supported by the STM observation of robust ZBPs near interstitial Fe adatoms \cite{yin2015observation,fan2021observation,kong2021majorana}. On the other hand, interactions of impurities with other relevant degrees of freedom may complicate the physical processes by generating features that could mask the origin of the ZBPs, thus requiring in-depth elucidation. Recent spin-resolved STM studies found equal weights between the opposite spin components of the ZBPs, raising the puzzling prospects that the ZBPs are of YSR-type, rather than the previously expected MZMs \cite{PhysRevLett.126.076802}.
Currently, interpretation of the measured STM spectra in systems with the coexistence of vortex and impurity are largely based on separate physics of impurity induced bound states and vortex induced bound states. However, the interaction between these distinct degrees of freedom may qualitatively change the benchmark properties. Therefore, understanding the interplay between magnetic impurity and vortex is crucial to clarifying the fundamental physics underlying the experimental observations.

\begin{figure}
    \centering
    \includegraphics[width=1.0\linewidth]{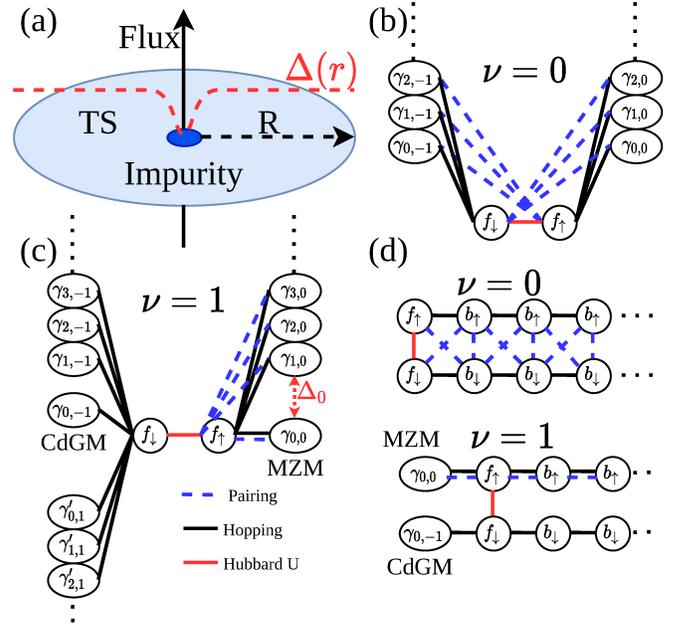}
    \caption{(a) Schematic plot of a magnetic flux pinned by a quantum magnetic impurity in a 2D TSC. (b, c) Effective star models for the TSC without or with a vortex, respectively. $\gamma_{n,m}$ denotes the Bogoliubov quasi-particles, where $n$ denotes the energy level and $m$ represents for the orbital angular momentum (OAM). The quasi-particles $\gamma^\prime_{n,1}$ are related to  $\gamma_{n,1}$ via the particle-hole transformation.
    (d) The effective chain model for the TSC without and with a vortex.}
    \label{fig:1}
\end{figure}

In this Letter, we study the generic problem of a vortex pinned by a quantum magnetic impurity, as illustrated in Fig.\ref{fig:1}(a), with the aim to unveil the salient effects of the quantum impurity on TSCs in the presence of the vortex. In stark contrast to the vortex free case, we find that for TSCs with a vortex where the MZM emerges, the previously established QPT between the singlet and doublet states disappears and is replaced by a crossover between a resonant scattering (RS) and a local moment (LM) regime. Moreover, both the impurity and the MZM generate low-energy peaks in the local density of states (LDOS) that possess opposite spin polarization. These results indicate that the interplay between the MZMs and Kondo effect can lead to ZBPs with comparable weights in distinct spin components, opening a new avenue for viable interpretation of the intriguing and puzzling experimental STM observations \cite{PhysRevLett.126.076802}.  These findings showcase emergent novel physics stemming from the interplay between MZMs and Kondo impurities, advancing fundamental understanding of the intricate physical processes and underlying mechanisms associated with the magnetic impurity pinned vortex in TSCs.

\textcolor{blue}{\textit{Impurity-vortex model in TSC}.--}
We consider a quantum magnetic impurity in 2D TSC with a magnetic flux pinned at the impurity as shown in Fig.\ref{fig:1}(a). This system is modeled by the Hamiltonian $H=H_{\textrm{imp}}+H_{\textrm{hyb}}+H_0$, where $H_{\textrm{imp}}$ describes an Anderson impurity
\begin{equation}\label{eqr1}
    H_{\textrm{imp}}=\sum_{\sigma} \epsilon_{f} \hat{f}_{\sigma}^{\dagger} \hat{f}_{\sigma}+U\hat{n}_{f\uparrow} \hat{n}_{f\downarrow},
\end{equation}
with $\hat{f}^\dagger_{\sigma}$ ($\hat{f}_{\sigma}$) being the creation (annihilation) operator for spin $\sigma$ of the impurity and $
\hat{n}_{f\sigma} = \hat{f}^\dagger_{\sigma} \hat{f}_{\sigma}$ the number operator. $\varepsilon_f$ is the energy of the local orbital and $U$ the on-site Hubbard interaction. We consider here the symmetric Anderson model with $\epsilon_f = -U/2$ and generalization to the asymmetric case is straightforward.

The local impurity is hybridized with the bath via
\begin{equation}\label{eqr2}
H_{\textrm{hyb}}=\sum_{\sigma} \int d \mathbf{r} V(r)\left[\hat{f}_{\sigma}^{\dagger} \hat{c}_{\mathbf{r}, \sigma}+h . c .\right],
\end{equation}
where $\hat{c}^\dagger_{\mathbf{r}, \sigma}$ ($\hat{c}_{\mathbf{r}, \sigma}$) is the creation (annihilation) operator of the SC bath at $\mathbf{r}$. We assume an isotropic coupling $V(\mathbf{r})=V(r)=V_0 e^{-r/r_0}$ and set $r_0 = 0.7 \xi$ \cite{PhysRevX.9.011033}, where $\xi$ is the superconducting coherence length and used as the length unit.

We consider a 2D TSC with effective $p+ip$ symmetry induced by spin-momentum locking, i.e,
\begin{multline}\label{eqr3}
H_0 = \int \mathrm{d}\mathbf{r}\;\sum_{\sigma,\sigma^\prime}\hat{c}^\dagger_{\mathbf{r},\sigma}\left[v_F\left[\left(\bm{\sigma}\times \mathbf{p}\right)\cdot\mathbf{z}\right]_{\sigma,\sigma^\prime}-\mu\delta_{\sigma,\sigma^\prime}\right]\hat{c}_{\mathbf{r},\sigma^\prime}\\
+\int \mathrm{d}\mathbf{r}\;\Delta(\mathbf{r})(\hat{c}^\dagger_{\mathbf{r,\uparrow}}\hat{c}^\dagger_{\mathbf{r,\downarrow}}+h.c.).
\end{multline}
Eq.\eqref{eqr1}-Eq.\eqref{eqr3} describe the Anderson impurity in extrinsic TSCs, e.g., the topological surface state in proximity to a normal $s$-wave SC or the surface TSC of $\mathrm{FeTe}_{0.55}\mathrm{Se}_{0.45}$\cite{PhysRevLett.117.047001, PhysRevX.9.011033}.

The magnetic flux pinned at the impurity generates the vortex state in TSCs, resulting in a winding phase in the gap function $\Delta(\mathbf{r})=\Delta(r)e^{i\nu\theta}$, with $\mathbf{r}=(r,\theta)$ and the winding number $\nu$ being an integer. We take $\Delta(r)=\Delta_0$ for $\nu=0$ and $\Delta(r)=r\Delta_0/\sqrt{r^2+\xi^2}$ for $\nu=1$\cite{PhysRevB.79.224506}, such that the characteristic length of the vortex is of the order of SC coherence length $\xi$. Although $\Delta(r)$ is not self-consistently determined here, it will not qualitatively affect the impurity states, as the specific form of $\Delta(r)$ only generates minor modifications to the CdGM vortex core states \cite{CAROLI1964307, PhysRevLett.124.097001}. In the following, we adopt the parameter values of $v_F=\sqrt{2\pi}V_0=1$, $\mu=0.2$ and $\Delta_0=0.1$ without losing generality in probing the main intended physics.

\textcolor{blue}{\textit{Mapping to star and chain models}.--}
The incorporation of a vortex into the quantum impurity problem is highly nontrivial. To take into account the contributions from the MZM and CdGM states in addition to the bulk electrons, we generalize the conventional mapping scheme in NRG calculations. Specifically, we adopt a disk geometry of the bath as shown in Fig.\ref{fig:1}(a). The resulting vortex-impurity model has rotational symmetry, allowing an expansion with respect to 2D spherical functions with the characteristic quantum number $m$, i.e., the orbital angular momentum (OAM). By diagonalizing $H_0$ through Bogoliubov transformation in real space, the original impurity model is mapped into the star models as shown in Fig.\ref{fig:1}(b) and Fig.\ref{fig:1}(c) for $\nu=0$ and $\nu=1$, respectively, with the impurity directly coupled to Bogoliubov quasi-particles\cite{supplemental}.

The OAM index $m$ plays a key role in determining the structure of the star model. First, for the vortex-free TSC ($\nu=0$), the impurity is coupled to the states with OAM $m=0$ and $m=-1$ \cite{PhysRevLett.122.087001}, as shown in Fig.\ref{fig:1}(b). These two orbitals form the pseudo-spin degree of freedom and screen the impurity. In the presence of a vortex ($\nu=1$), however, the situation is totally different, as shown in Fig.\ref{fig:1}(c). The spin-down impurity state couples to the Bogoliubov particles and CdGM modes via hopping only, whereas the spin-up state couples to the Bogoliubov particles and the MZM via both hopping and pairing.  The different coupling mechanisms can lead to distinct impurity states with the presence of a vortex.

We further tridiagonalize the star models by the block Lanczos approach \cite{PhysRevB.93.035102,PhysRevB.90.195109,allerdt_numerically_2019, supplemental} to generate 1D chains with only nearest neighbor interactions, where the impurity is located at the boundary, as shown in Fig.\ref{fig:1}(d). We preserve the MZM-impurity and CdGM-impurity couplings in this procedure to keep track of their distinct impacts on the system. The chain length depends on the truncation of the Bessel basis $j_\mathrm{max}$. We use $j_\mathrm{max}=100$ in our calculations. The resulting chain is of the length $400$ ($200$) for the spin-down (spin-up) environment.

The star models in Fig.\ref{fig:1}(b)-(c) and chain models in Fig.\ref{fig:1}(d) exhibit different symmetries for TSC without or with vortex. The real spin in the TSC is not conserved because of the spin-momentum locking. However, the pseudo-spin $S_z$, which comprises $m=0$ and $m=-1$, is a conserved quality for $\nu=0$. However, the vortex further breaks the pseudo-spin symmetry, leaving only two conserved quantities, namely the total number $N_{\downarrow}$ of the quasi-particles coupled to $f_{\downarrow}$, and the  number parity $P_\uparrow$ of the quasi-particles coupled to $f_{\uparrow}$.

\begin{figure}
    \centering
    \includegraphics[width=1.0\linewidth]{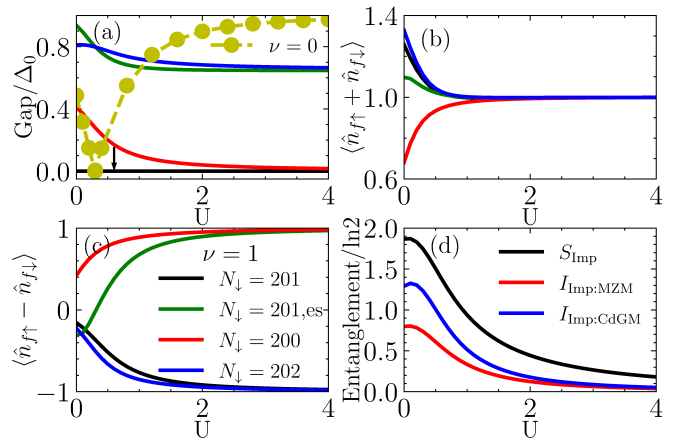}
    \caption{(a) Calculated energy gap $\Delta E_i$ as a function of $U$ for $\nu=1$, which indicates the occurrence of a number of in-gap states. The states shown are all doubly degenerate for even and odd $P_\uparrow$, as a result of the MZM. For comparison, the result for $\nu=0$ case is shown by the yellow curve, which displays a singlet-doublet QPT. The occupation number (b)  and the magnetic momentum (c) obtained at the impurity versus $U$ for $\nu=1$. The same legends are used in (a-c) to denote the states in different symmetry sectors.  ``es'' denotes the first excited state in the corresponding sector. (d) shows the entanglement entropy for the impurity, and the mutual information between the impurity and MZM/CdGM.}
    \label{fig:2}
\end{figure}

\textcolor{blue}{\textit{Unique in-gap states driven by MZM and CdGM}.--}
We use DMRG to target all in-gap states for the chain models in Fig.\ref{fig:1}(d) \cite{supplemental}. In Fig.\ref{fig:2}(a) we show energy gaps defined by $\Delta E_i=E_i-E_0$, where $E_i$ and $E_0$ are the energies of the $i$-th excited states and ground state, respectively. For the vortex-free case $\nu=0$, $\Delta E_1$ firstly decreases to zero and then increases with raising $U$ \cite{yoshioka2000numerical, PhysRevLett.122.087001, RevModPhys.78.373}, as shown by the yellow curve in Fig.\ref{fig:2}(a). The critical point $U_c$ indicates a QPT, separating the singlet ($U<U_c$) and doublet ($U>U_c$) ground state \cite{PhysRevLett.122.087001}.

We now focus on TSC with a vortex. Utilizing the conserved quantities $P_{\uparrow}$ and $N_{\downarrow}$, all the low-energy states can be classified into different sectors.  In each sector, we calculate the low-energy spectrum,  the average occupation number at the impurity  $\langle \hat{n}_{f\uparrow}+\hat{n}_{f\downarrow}\rangle$ and the impurity   moment $\langle \hat{n}_{f\uparrow}-\hat{n}_{f\downarrow}\rangle$, which are shown in Fig.\ref{fig:2}(a), (b) and (c), respectively. As shown in Fig.\ref{fig:2}(a), the previously established singlet-doublet transition for the vortex free case vanishes. Instead, a number of in-gap states take place, without showing any signature of a QPT.

The impurity ground state can be understood at the two limits, i.e., $U=0$ and $U\to \infty$. For $U=0$, the charge fluctuations become dominant,  as we find in Fig.2(c) that $\langle \hat{n}_{f\uparrow}-\hat{n}_{f\downarrow} \rangle \sim 0$ for $U=0$, which implies that the ground state is located in the resonant scattering (RS) regime \cite{yoshioka2000numerical,supplemental2}. For $U\to \infty$, the large Hubbard $U$ ensures the single-occupancy condition at the impurity, as electron number $\langle \hat{n}_{f\uparrow}+\hat{n}_{f\downarrow}\rangle$ approaches 1 for large $U$, as shown by Fig.\ref{fig:2}(b). Moreover, as indicated by Fig.\ref{fig:2}(c), the local moment $\langle \hat{n}_{f\uparrow}-\hat{n}_{f\downarrow}\rangle$ saturates to $\pm1$ for large $U$, showing that the impurity remains effectively unscreened and a doublet ground state emerge at this limit. With increasing $U$ from 0 to $\infty$, there is no signature of any QPTs, implying a crossover from the RS  to the LM regime.

To further validate the nature of ground state, we also calculate the entanglement entropies (EE) $S_A$, which measures the entanglement between subsystem A and its environment. The black curve in Fig.\ref{fig:2}(d) is the result of $S_{\textrm{Imp}}$. $S_{\textrm{Imp}}\rightarrow2\ln 2$ is reached for $U\rightarrow0$, indicating that not only the spin but also the charge degrees of freedom are entangled with the bath. $S_\textrm{Imp}$ decreases down to vanishingly small values with increasing $U$, reflecting a LM regime with effectively free local moment. We also calculated the mutual information\cite{PhysRevLett.122.067203} between the impurity and MZM or CdGM state. The mutual information is defined as $I_{\textrm{A}:\textrm{B}} = S_\textrm{A} + S_\textrm{B} - S_\textrm{AB}$, which measures the entanglement between two subsystems A and B. As shown in Fig.\ref{fig:2}(d), both $I_{\textrm{Imp}:\textrm{MZM}}$ and $I_{\textrm{Imp}:\textrm{CdGM}}$ vanish at large $U$, which indicates that the correlations between the impurity and the in-gap modes are indeed negligible in the LM regime.

\begin{figure}
  \centering
  \includegraphics[width=1.0\linewidth]{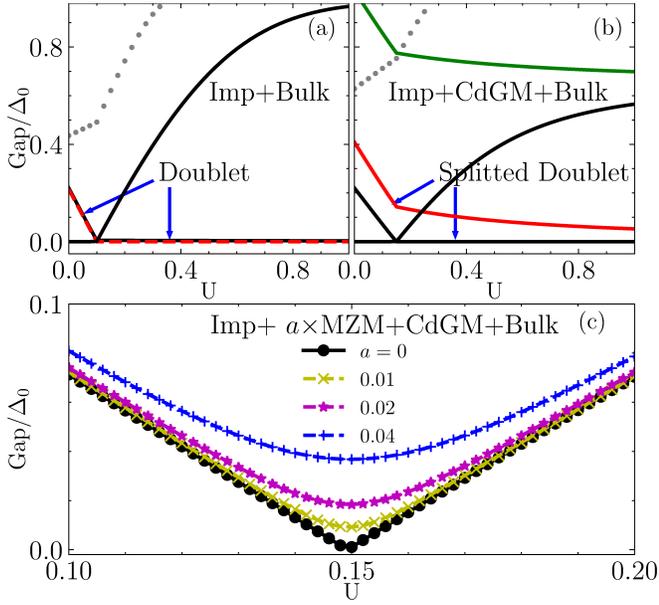}
  \caption{The in-gap states as a function of $U$ for different couplings to the impurity. (a) shows the spectrum with turning off the impurity-MZM and impurity-CdGM coupling, and (b) shows the results with turning off the impurity-MZM coupling. (c) The zoom-in spectrum near the critical point with gradually turning on the impurity-MZM coupling controlled by the factor $a$. After completely turning on the impurity-MZM coupling with $a=1$, the green (red, blue) curve in Fig.\ref{fig:2}(a) will be reproduced by the continuous evolution of the green (red, blue) curve in (b) and (c), while the dotted curve in (a) or (b) will be pushed into the continuum.}
  \label{fig:3}
\end{figure}

To examine the crossover behavior in the intermediate $U$ regime, we divide the system into four parts: the impurity, the CdGM state, the MZM state, and the bulk states consisting of the Bogoliubov particles with $E>\Delta_0$. Then, we evaluate the low-energy spectrum with artificially tuning the impurity-CdGM and impurity-MZM coupling. First, we turn off both the impurity-CdGM and impurity-MZM coupling. The spectrum shown in Fig.\ref{fig:3}(a) well reproduces the vortex-free case in Fig.\ref{fig:2}(a), displaying a singlet-doublet QPT. Then, we turn on the impurity-CdGM coupling. Interestingly,  although the feature of QPT is still preserved, the CdGM induces a significant splitting of the doublet state, as shown by Fig.\ref{fig:3}(b). In fact, through a Schrieffer-Wolff transformation \cite{PhysRevB.37.9312}, we show that the impurity-CdGM coupling essentially contributes to an effective Zeeman field to the impurity \cite{supplemental}.

We then gradually turn on the impurity-MZM coupling controlled by a factor $a$ in front of the coupling. Fig.\ref{fig:3}(c) shows the spectrum around the critical point $U_c$ with raising $a$ up to $0.04$. As long as $a$ is nonzero, the gapless point of the QPT is turned into an avoided level crossing, reproducing the previously observed crossover. The avoided level crossing is a direct result of the model symmetry. With $a=0$, the parity of impurity-bulk system  $P^\textrm{Bulk+Imp}_\uparrow$ is conserved. The even and odd parity sectors are degenerate at the critical point. However, such a degeneracy is broken as long as the MZM-impurity coupling is turned on, which in turn gaps out the critical point \cite{supplemental}.
 
\begin{figure}
    \centering
    \includegraphics[width=1.0\linewidth]{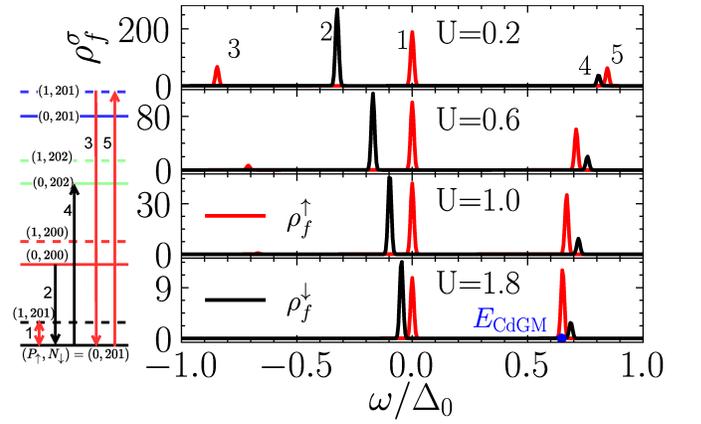}
    \caption{Spin-resolved in-gap LDOS $\rho_f^{\sigma}$ for different values of $U$. The left corner builds up the one-to-one correspondence between the LDOS peaks  and the level transition processes (both are labeled by the numbers from 1 to 5).  $(P_\uparrow, N_\downarrow)$ denotes the symmetry sectors introduced in Fig.\ref{fig:2}. Dotted and solid levels with the same color are degenerate in energy, which are shifted intentionally for the purpose of clarity.}
    \label{fig:4}
\end{figure}

\textcolor{blue}{\textit{Spin-resolved LDOS}.--}
We now calculate the spin-resolved local density of states (LDOS) at the impurity site under zero temperature, which is crucial to understanding the STM experiments \cite{PhysRevLett.119.197002, PhysRevB.82.075111, PhysRevLett.92.256602, PhysRevB.82.075111}. In the Lehmann representation
\begin{multline}
\rho_f^{\sigma}(\omega) = \sum_{m}\left[\left|\langle \psi_m | f_{\sigma}^{\dagger}|\psi_0\rangle\right|^2 \delta(\omega +E_0 - E_m)\right.\\ + \left.\left|\langle \psi_0 | f_{\sigma}^{\dagger}|\psi_m\rangle\right|^2 \delta(\omega + E_m - E_0)\right],
\end{multline}
where $E_0$ ($E_m$) is the energy of the ground state $|\psi_0\rangle$ ($m$-th excited state $|\psi_m\rangle$). The LDOS is then obtained with taking into account all the in-gap states as shown by Fig.\ref{fig:2}(a), whose energies satisfy $E_m-E_0<\Delta$. Besides, a broadening factor $b=0.001$ is chosen in the calculation to show the peaks more clearly.

As shown by Fig.\ref{fig:4}, the LDOS is highly spin sensitive, which is expected since the TSC bath is effectively  spin polarized. For $\rho_f^\uparrow$, there always takes place a robust zero-energy peak \cite{PhysRevB.89.165314,PhysRevB.91.115435}, as a result of the MZM, which is only coupled to $f_{\uparrow}$ as indicated by Fig.\ref{fig:1}(d). Besides, two more peaks arise in $\rho_f^\uparrow$ at higher energy; one comes from the hole component ($\omega<0$) and the other from the electron component ($\omega>0$). With increasing $U$, both peaks gradually move toward $\omega=0$. Besides, the hole component peak disappears around $U\sim 1$, while the electron component saturates at the energy of the CdGM, i.e., $E_{\textrm{CdGM}}$ for $U\gtrsim 1$, as indicated by Fig.\ref{fig:4}.

Remarkably, distinct LDOS feature is found for $\rho_f^\downarrow$. A significant peak (labeled by ``2") emerges, which is originated from the transition between the ground state the first excited state,as indicated by Fig.\ref{fig:2}(a). Hence, it is a direct consequence of the RS-LM crossover. With increasing $U$, the peak continuously evolves and approaches to $\omega=0$ at large $U$. Furthermore, for $U\gtrsim1.8$, the peak enjoys comparable height but opposite spin-polarization as that of the MZM peak, signifying an interesting LDOS feature relevant to experiments.  Last, we also clarify the origin of all the LDOS peaks by showing their one-to-one correspondence with the level transition processes, as indicated by Fig.\ref{fig:4}.

It should be noted that the occurrence of two LDOS peaks with comparable weight and opposite spin-polarization near zero energy has been observed in recent experiment in TSC \cite{PhysRevLett.119.197002}. Although it was explained by YSR states, our findings show new possibilities stemming from the novel crossover associated with the  impurity-pinned vortex state.

\textcolor{blue}{\textit{Conclusion and discussion}.-- }
We have  employed the DMRG method to study magnetic impurity pinned vortex in TSC. The results reveal fundamentally distinct behaviors of the in-gap states in TSC, which are driven by three physical processes in cooperation with each other. First, the impurity-bulk coupling generates the competition between Kondo screening and Cooper pairing. Second, the impurity-CdGM coupling leads to an effective Zeeman splitting for the impurity. Third, the impurity-MZM coupling produces a crossover between the RS and LM regime with increasing the Hubbard $U$. The crossover behavior as well as the in-gap states are expected to be general properties of the impurity-pinned vortex state in superconductors due to the following reasons. First, the crossover is the result of the model symmetries at the critical point, and it should occur as long as the impurity-MZM coupling breaks the conservation of total spin (pseudospin).  Second,  since the TSC bath is  polarized by the vortex, the in-gap states are robust against spin-dependent perturbation, such as the external magnetic field or asymmetric parameters in the Anderson model. Remarkably, over a wide parameter range, we show that, in addition to the MZM, the crossover generates another LDOS peak near zero-energy, which exhibits the comparable weight but opposite spin polarization as that of MZM. The LDOS and the in-gap physics found here are originated from the mutual effects from the CdGM and MZM in TSC, and are clearly different in nature from all known impurity states in superconductors.  Our results clarify the long-standing impurity problem in TSCs with a vortex, which could provide further guidance to the experimental studies of TSCs.

\begin{acknowledgments}
We are grateful to Lin Li and Shang Xin for fruitful discussions. This work was supported by the Natural Science Foundation of China (No. 11904245) and Postdoctoral Science Foundation of China (No. 2021M690330), and was also supported by the High Performance Computing Center of Sichuan Normal University, China.

\end{acknowledgments}

\clearpage

\appendix

\widetext
\begin{center}
	\textbf{\large Supplementary Material for: Unique quantum impurity states driven by a vortex in topological superconductors}
\end{center}

\vspace{8mm}
\renewcommand\thefigure{\thesection S\arabic{figure}}
\setcounter{figure}{0}
\renewcommand\theequation{\thesection S\arabic{equation}}
\setcounter{equation}{0} 

\twocolumngrid

\section{An outline of the methods}
The incorporation of an additional vortex into the quantum impurity problem is highly nontrivial at least in the following two senses. First, the DOS of the bath is complicated by the CdGM vortex core states and MZM, thus the conventional numerical renormalization group (NRG) method cannot be directly applied. Second, the interplay between the quantum magnetic impurity and the MZMs can sharply alter the ground state of impurity state, resulting in novel physical features available by STM measurements. We adopted the numerical scheme specified by the following steps, using which we can accurately extract the effect of the bulk states as well as the in-gap bound states on the impurity physics.
\begin{enumerate}
  \item Diagonalization of the TSC via Bogoliubov transformation in real space. The impurity is then explicitly coupled to the Bogoliubov quasi-particles, which is referred to the star model in the main text. In this way we can directly keep track of the impurity-MZM and impurity-CdGM coupling (see Sec.\ref{appendix1} for details.).
  \item Tridiagonalization of the star model into a chain with only the nearest neighbor interactions through block Lanczos. The process can preserve the form of the impurity-MZM and impurity-CdGM couplings (see Sec.\ref{appendix2} for details.).
  \item Application of the DMRG solver to obtain all the in-gap states of the chain model (see Sec.\ref{appendix3} for details on implementation and assurance of convergence).
\end{enumerate}
No approximation is made within the above scheme, except for (1) a truncation of the Bessel functions in step 1 and (2) truncation of basis in DMRG. The  error introduced by both truncations are controllable by examining the computational convergence.

\section{ \label{appendix1} Bogoliubov transformation}
In Nambu basis $\Psi(\mathbf{r}) = (c_{\mathbf{r},\uparrow},c_{\mathbf{r},\downarrow},c^\dagger_{\mathbf{r},\downarrow},-c^\dagger_{\mathbf{r},\uparrow})^{T}$, $H_0 = \int\mathrm{d}\mathbf{r}\;\Psi^\dagger(\mathbf{r}) \mathcal{H}_0 \Psi(\mathbf{r})$.  In terms of Nambu spinor $\Phi_n(\mathbf{r}) \equiv (u_{n\uparrow},u_{n\downarrow},v_{n\downarrow},-v_{n\uparrow})^T$, the Bogolubov-de Gennes (BdG) equation is $\mathcal{H}_0 \Psi_n(\mathbf{r}) = E_{n}\Psi_n(\mathbf{r})$, with
\begin{equation}
\mathcal{H}_0 =
\left(
\begin{array}{cc}
H & \Delta (\mathbf{r}) \\
\Delta^\ast(\mathbf{r}) & -\sigma_y H^{\ast} \sigma_y
\end{array}
\right).
\end{equation}
The Hamiltonian $\mathcal{H}_0$ can be diagonalized by Bogoliubov transformation,
\begin{eqnarray}
c_{\sigma}(\mathbf{r}) &=& \sum_{n}\left[u_{n\sigma}(\mathbf{r})\gamma_{n}+v_{n\sigma}^{\ast}(\mathbf{r})\gamma_{n}^{\dagger}\right],
\end{eqnarray}
\begin{eqnarray}
c^\dagger_{\sigma}(\mathbf{r}) &=& \sum_{n}\left[u^\ast_{n\sigma}(\mathbf{r})\gamma^\dagger_{n}+v_{n\sigma}(\mathbf{r})\gamma_{n}\right].
\end{eqnarray}
Note that particle-hole symmetry ensures that if $\Phi_{n} = \left[ u_{n,\uparrow}, u_{n,\downarrow}, v_{n,\downarrow}, -v_{n\uparrow} \right]^{T}$ is an eigenstate with energy $E_n$, then $\left[ -v^\ast_{n,\uparrow}, -v^\ast_{n,\downarrow}, -u^\ast_{n,\downarrow}, u^\ast_{n\uparrow} \right]^{T}$ is also an eigenstate with energy $-E_n$. After transformation,
\begin{equation}
\label{eq:eg}
H_0 = \sum_{n} E_{n} \gamma_{n}^{\dagger}\gamma_{n} + E_g,
\end{equation}
where $E_g$ is the ground state energy and, the summation is over the excitation states with positive energy.

The inverse transformation is given by
\begin{subequations}
\begin{equation}
\gamma_n^{\dagger} = \int\mathrm{d}\mathbf{r}\;\sum_{\sigma}\left[u_{n\sigma}(\mathbf{r})c_{\sigma}^\dagger(\mathbf{r})+v_{n\sigma}(\mathbf{r})c_{\sigma}(\mathbf{r})\right],
\end{equation}
\begin{equation}
\gamma_n = \int\mathrm{d}\mathbf{r}\;\sum_{\sigma}\left[u^\ast_{n\sigma}(\mathbf{r})c_{\sigma}(\mathbf{r})+v^\ast_{n\sigma}(\mathbf{r})c^\dagger_{\sigma}(\mathbf{r})\right],
\end{equation}
\end{subequations}
with $\gamma_n^\dagger$ and $\gamma_n$ being the creation and destruction operators of Bogoliubov quasi-particles. They satisfy the fermionic anti-commutation relation
\begin{eqnarray}
    \{\gamma_{n},\gamma_{n^\prime}\}=\{\gamma_{n}^{\dagger},\gamma_{n^\prime}^{\dagger}\} = 0, \ \ \{\gamma_{n},\gamma_{n^\prime}^{\dagger}\} = \delta_{nn^\prime}.
\end{eqnarray}

\begin{figure}
    \centering
    \includegraphics[width=0.95\linewidth]{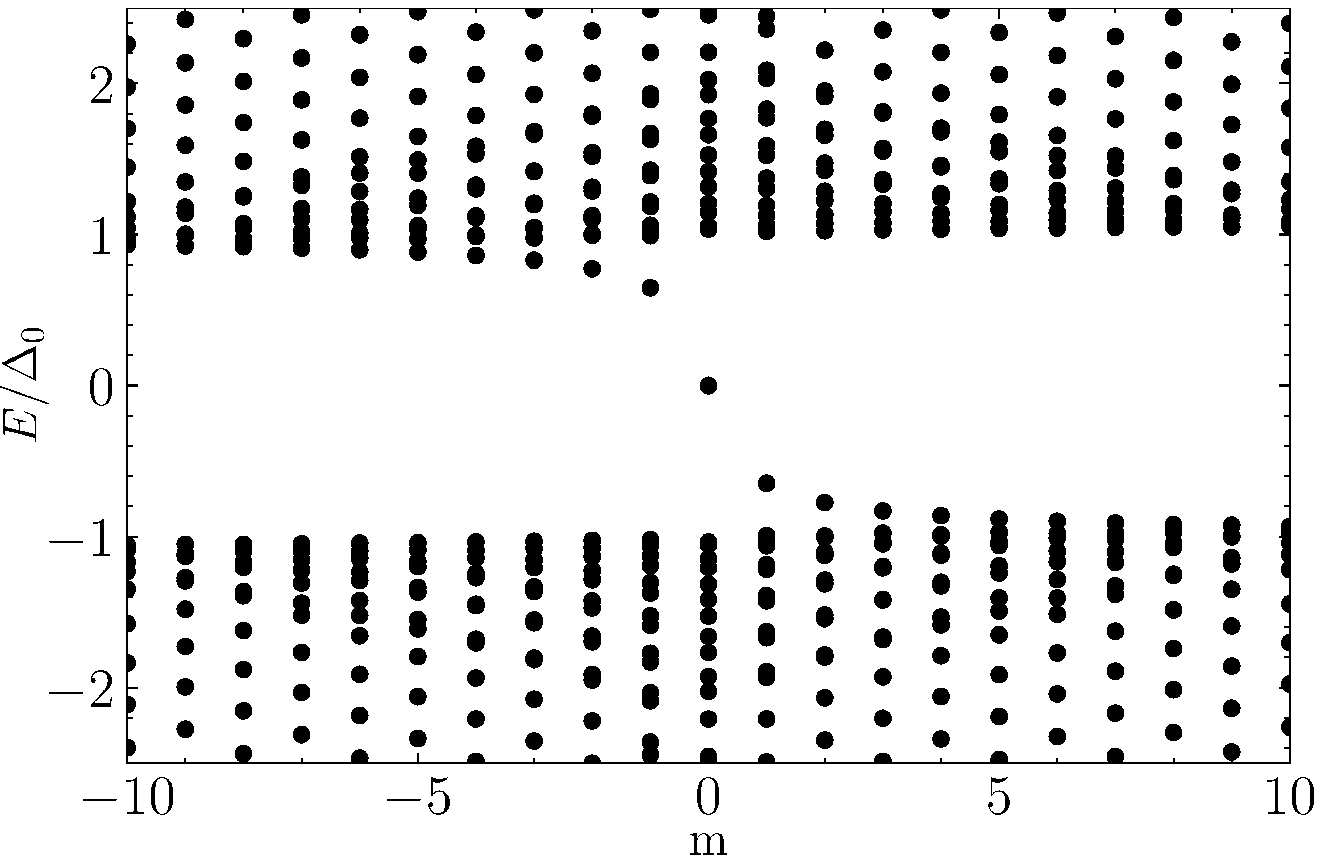}
    \caption{Single particle spectrum for $\nu=1$ without impurity. $m$ is the OAM index.}
    \label{fig:A1}
\end{figure}

For the TSC, the helical Dirac surface state carries an extra Berry phase as $\partial_x\pm \partial_y = \exp(\pm i\theta) \left(\partial_r \pm \frac{i}{r}\partial_\theta\right)$\cite{PhysRevLett.107.097001, PhysRevX.9.011033}. The BdG equation thus can be solved by assuming the following wave function
\begin{multline}
\Phi^{\textrm{TSC}}_{nm}(r,\theta) = e^{im\theta}\left[ u_{nm\uparrow}(r), u_{nm+1\downarrow}(r)e^{i\theta},\right.\\
\left. v_{nm-\nu\downarrow}(r)e^{-i\nu\theta}, -v_{nm-\nu+1\uparrow}(r)e^{-i(\nu-1)\theta}\right]^{T},
\end{multline}
where $m = 0,\pm 1, \pm 2,\cdots$ are integers. The particle-hole symmetry is defined between OAM $m$ and $-m+\nu-1$. For $\mu=0$, there exists another particle-hole symmetry within each OAM $m$. This symmetry is inherited from the particle-hole symmetry of the surface state at the Dirac point.

The Bogoliubov transformation that diagonalizes $\mathcal{H}_0$ of the TSC is given by
\begin{multline}
c_{\uparrow}(\mathbf{r})=\frac{1}{\sqrt{2\pi}}\sum_{nm} \left[e^{i m \theta } u_{nm\uparrow}(r) \gamma_{nm}\right.\\
\left. + e^{-i m\theta } v^\ast_{nm-\nu+1\uparrow}(r)e^{i(\nu-1)\theta}\gamma_{nm}^{\dagger} \right],
\end{multline}
\begin{multline}
c_{\downarrow}(\mathbf{r})=\frac{1}{\sqrt{2\pi}}\sum_{nm} \left[e^{i m \theta } u_{nm+1\downarrow}(r)e^{i\theta} \gamma_{nm}\right.\\
\left. + e^{-i m \theta } v^\ast_{nm-\nu\downarrow}(r)e^{i\nu\theta}\gamma_{nm}^{\dagger} \right].
\end{multline}
In practice, the radial functions $u(r)$ and $v(r)$ are solved by expanding with the Bessel functions
\begin{equation}
\phi_{m,j}(r) = \frac{\sqrt{2}}{RJ_{m+1}(\beta_{m,j})}J_m\left(\beta_{m,j}\frac{r}{R}\right).
\end{equation}
These Bessel functions form a complete set of basis on a disc with radius $R$ and $j=1, 2,\cdots, j_{\textrm{max}}$. $j_{\textrm{max}}$ should be $\infty$ in principle, but a truncation is used in practice.
The BdG equation is then solved in the basis of
\[
(u_{m,j,\uparrow},u_{m+1,j,\downarrow},v_{m-\nu,j,\downarrow},-v_{m-\nu+1,j,\uparrow})_{j=1\cdots j_{\textrm{max}}}.
\]
Diagonalizing the corresponding $4j_{\textrm{max}}\times 4j_{\textrm{max}}$ Hamiltonian matrices numerically, we get the Bogoliubov quasi-particle spectrum in Fig.\ref{fig:A1}. $m$ being an integer for the TSC with a vortex is crucial for supporting the MZM in the OAM $m=0$\cite{PhysRevLett.107.097001}. In addition to the MZM, there are CdGM states\cite{caroli_bound_1964}, carrying integer quantum numbers of $m=\pm 1,\pm 2,\cdots$.

After transformation, the hybridization between the impurity and TSC is
\begin{multline}
\label{eq:hyb2}
H^{\textrm{TSC}}_{\textrm{hyb}} = \sum_{\sigma}\int \mathrm{d}\mathbf{r}\; V(r)\hat{f}_{\sigma}^\dagger\hat{c}_{\mathbf{r},\sigma}+h.c.\\
= \sqrt{2\pi}\sum_{n}\int \mathrm{d}r\;r V(r) f^{\dagger}_{\uparrow} \left[u_{n0\uparrow}(r)\gamma_{n0} + v^\ast_{n0\uparrow}(r)\gamma_{n\nu-1}^{\dagger}\right]\\
+ \sqrt{2\pi}\sum_{n}\int \mathrm{d}r\;r V(r) f^{\dagger}_{\downarrow} \left[u_{n0\downarrow}(r)\gamma_{n,-1} + v^\ast_{n0\downarrow}(r)\gamma_{n,\nu}^{\dagger}\right]\\ + h.c..
\end{multline}
It is seen that both $f_{\uparrow}$ and $f_{\downarrow}$ couple to $m=0,-1$ channels for $\nu=0$, while $f_{\uparrow}$($f_{\downarrow}$) couples to the $m=0$($m=\pm 1$) channel(s) for $\nu=1$.

Eq.\eqref{eq:hyb2} combined with Eq.\eqref{eq:eg} and $H_{\textrm{imp}}$ defines the star models, where the impurity directly couples to Bogoliubov quasi-particles. For TSC with $\nu=1$, a particle-hole transformation can be performed for $\gamma^\dagger_{n, 1}\to \gamma^\prime_{n, 1}$. The pairings between the impurity and $\gamma^\dagger_{n, 1}$ are transformed  into hopping terms between the impurity and $\gamma^\prime_{n, 1}$, correspondingly. Note that the pairing persists in the coupling between $f_\uparrow$ and $\gamma_{n, 0}$.

\section{ \label{appendix2} Block Lanczos and the chain models}
We use block Lanczos to transform the star models to chain models that contain only nearest interactions\cite{PhysRevB.93.035102,PhysRevB.90.195109,allerdt_numerically_2019}, which are in a more suitable form for numerical calculations. Using the coupling coefficient in Eq.\eqref{eq:hyb2} as the initial vector and tridiagonalize the bulk Hamiltonian Eq.\eqref{eq:eg}, the resulting chain model for the vortex-free TSC is obtained as
\begin{equation}
H_{\textrm{chain}}^{\nu=0} = H_\textrm{imp}+\left(\sum_{\sigma}t_{0\sigma} f^\dagger_{\sigma} b_{1\sigma} + \Delta^\prime_{0} f_{\downarrow} b_{1\uparrow}+h.c.\right) + H_{\textrm{env}}^{\nu=0}
\end{equation}
where
\begin{multline}
H^{\nu=0}_{\textrm{env}}=\sum_{\sigma, i=1}^{2j_{\textrm{max}}} \varepsilon_{i\sigma} b^{\dagger}_{i\sigma}b_{i\sigma} + \sum_{i=1}^{2j_{\textrm{max}}} \left(\Delta_{i} b^{\dagger}_{i\uparrow}b^\dagger_{i\downarrow}+h.c.\right)\\
 + \sum_{i=1}^{2j_{\textrm{max}}-1}\left(\sum_{\sigma}t_{i\sigma} b^\dagger_{i\sigma} b_{i+1\sigma} + \Delta^\prime_{i} b_{i\downarrow} b_{i+1\uparrow} + h.c.\right)
\end{multline}

For $\nu=1$, it is desirable to leave the form of the impurity-MZM ($\gamma_{0,0}$) and impurity-CdGM ($\gamma_{0,-1}$) coupling intact. This can be achieved by making the coupling terms between $f_\sigma$ and $\gamma_{0,0(-1)}$  in Eq.\eqref{eq:hyb2} invariant, while using other coupling terms as the initial vector to tridiagonalize the bulk Hamiltonian, $H_{0}^\prime = \sum_{n,m=0,\pm 1}^\prime E_{n,m} \gamma_{n,m}^\dagger\gamma_{n,m}$ (Here $\sum_{n,m=0,\pm 1}^\prime$ excludes terms with $(n,m) = (0,0)$ and $(0,-1)$). This procedure leads to the following chain model:
\begin{multline}
  H_{\textrm{chain}}^{\nu=1} = H_{\textrm{imp}} + H_{\textrm{chain}}^{\nu=1,\uparrow} + H_{\textrm{chain}}^{\nu=1,\downarrow}\\
   + E_{0,0}\gamma_{0,0}^\dagger\gamma_{0,0} + E_{0,-1}\gamma_{0,-1}^\dagger\gamma_{0,-1}\\
   +[f^\dagger_{\uparrow}(U_{00\uparrow}\gamma_{0,0} + V_{00\uparrow}\gamma_{0,0}^\dagger) +
  U_{00\downarrow}f^\dagger_{\uparrow}\gamma_{0,-1}+h.c.]
\end{multline}
where $U_{nm\sigma} = \sqrt{2\pi}\int \mathrm{d}r\; r V(r)u_{nm\sigma}(r)$, $V_{nm\sigma} = \sqrt{2\pi}\int \mathrm{d}r\;r V(r)v^\ast_{nm\sigma}(r)$, and
\begin{equation}
  H^{\nu=1,\uparrow}_{\textrm{chain}} = \left(t_0 f^\dagger_{\uparrow} b_{1\uparrow} + \Delta_0 f_{\uparrow} b_{1\uparrow} +h.c. \right)+ H^{,\nu=1,\uparrow}_{\textrm{env}},
  \end{equation}
\begin{multline}
H^{\nu=1,\uparrow}_{\textrm{env}}=\sum_{i=1}^{2j_{\textrm{max}}} \varepsilon_i b^{\dagger}_{i\uparrow}b_{i\uparrow} \\
+\sum_{i=1}^{2j_{\textrm{max}}-1}\left(t_i b^\dagger_{i\uparrow} b_{i+1\uparrow} + \Delta_i b_{i\uparrow} b_{i+1\uparrow}+h.c.\right),
\end{multline}
\begin{equation}
H^{\nu=1,\downarrow}_{\textrm{chain}} = \left(t_0 f^\dagger_{\downarrow} b_{1\downarrow} +h.c. \right)+ H^{\nu=1,\downarrow}_{\textrm{env}},
\end{equation}
\begin{equation}
H^{\nu=1,\downarrow}_{\textrm{env}}=\sum_{i=1}^{4j_{\textrm{max}}} \varepsilon_i b^{\dagger}_{i\downarrow}b_{i\downarrow} +\sum_{i=1}^{4j_{\textrm{max}}-1}\left(t_i b^\dagger_{i\downarrow} b_{i+1\downarrow} + h.c.\right)
\end{equation}

\section{ \label{appendix3} Numerical details of DMRG calculations}
The largest number of states kept in DMRG is $2,000$ for TSC without the vortex, with the truncation error on the density matrix $\epsilon < 10^{-10}$. For the TSC with $\nu = 1$, keeping about $300$ states is sufficient to reach the same level of accuracy. For $U=0$, the resulting errors in energy are less than $10^{-7}$ compared to the exact results. We used $R=100$ and $j_\textrm{max}=100$ in our calculations, with the topological degeneracy preserved to the order of $10^{-3}\Delta_0$. The convergence of $j_\textrm{max}$ is carefully checked with $j_\textrm{max}= 60\to 140$, with energy differences being less than $10^{-3}\Delta_0$.

We mention here that although it is easier to obtain the ground state and the low-energy excited states in the gapped TSC compared to the gapless thermal baths, local minimums can still take place if one starts from random initial states. To avoid the local minimums, we first obtain the low-energy states for short chains (with the impurity), and then successively add into the system more and more sites of the environment.

\section{Schrieffer-Wolff transformation}
To reveal more clearly the effect of the CdGM mode, we perform the Schrieffer-Wolff transformation $\mathcal{U} = e^{\mathcal{S}}$ to eliminate the coupling term to linear order in $V$\cite{PhysRevB.37.9312}.
The unitarity requires $\mathcal{U}^{-1} = \mathcal{U}^\dagger$, implying $\mathcal{S}^\dagger = -\mathcal{S}$. The transformed Hamiltonian can be expressed as follows:
\begin{eqnarray}
  H_{\textrm{eff}} &=& \mathcal{U} H \mathcal{U}^\dagger = \sum_{0}^{\infty} \frac{1}{n!} [\mathcal{S}, H]_n \nonumber\\
  & \approx & \mathcal{H}_0 + H_{\textrm{hyb}} + [\mathcal{S}, \mathcal{H}_0] + [\mathcal{S}, H_{\textrm{hyb}}] + \frac{1}{2}[[\mathcal{S}, [\mathcal{S}, \mathcal{H}_0]]\nonumber\\
\end{eqnarray}
with $\mathcal{H}_0 \equiv H_0 + H_{\textrm{imp}}$. By demanding
\begin{equation}
  H_{\textrm{hyb}} + [\mathcal{S}, \mathcal{H}_0] = 0,
\end{equation}
$H_{\textrm{hyb}}$ is then eliminated to linear order, and the transformed Hamiltonian becomes
\begin{equation}
  H_{\textrm{eff}} = \mathcal{H}_0 + \frac{1}{2}[\mathcal{S}, H_{\textrm{hyb}}].
\end{equation}

\begin{figure}
  \centering
  \includegraphics[width=0.95\linewidth]{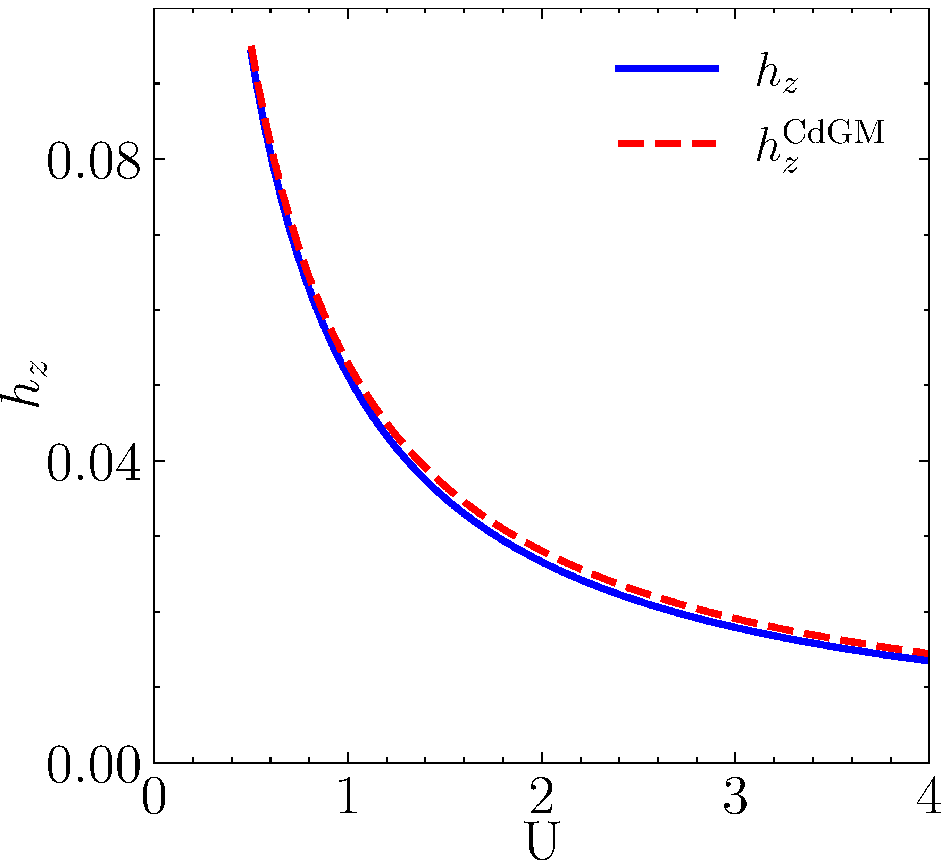}
  \caption{Strength of the effective Zeeman term and that contributed by the CdGM mode as a function of $U$.}
  \label{fig:A2}
\end{figure}
The generator for $\mathcal{S} = \mathcal{S}_0 - \mathcal{S}_0^\dagger$ reads as
\begin{multline}
  \mathcal{S}_0 = \sum_n \left\{
    U^\ast_{n\uparrow} \left[ \frac{1-n_{f\downarrow}}{E_{n,0} - \epsilon_{\uparrow} } + \frac{n_{f\downarrow}}{E_{n,0} - \epsilon_{\uparrow} - U }\right] \gamma_{n,0}^\dagger f_{\uparrow}\right. \\
    - V^\ast_{n\uparrow} \left[ \frac{1-n_{f\downarrow}}{E_{n,0} + \epsilon_{\uparrow} } + \frac{n_{f\downarrow}}{E_{n,0} + \epsilon_{\uparrow} + U }\right] \gamma_{n,0} f_{\uparrow} \\
    + U^\ast_{n\downarrow} \left[ \frac{1-n_{f\uparrow}}{E_{n,-1} - \epsilon_{\downarrow} } + \frac{n_{f\uparrow}}{E_{n,-1} - \epsilon_{\downarrow} - U }\right] \gamma_{n,-1}^\dagger f_{\downarrow}\\
    -\left. V^\ast_{n\downarrow} \left[ \frac{1-n_{f\uparrow}}{E_{n,1} + \epsilon_{\downarrow} } + \frac{n_{f\uparrow}}{E_{n,1} + \epsilon_{\downarrow} + U }\right] \gamma_{n,1} f_{\downarrow}
  \right\},\\
\end{multline}
which can be checked by using $[n, f] = -f$ and $[n, f^\dagger] = f^\dagger$.
The transformed Hamiltonian can be arranged into a concise form
\begin{multline}
  H_{\textrm{eff}} = \mathcal{H}_0^\prime + \mathcal{H}_\textrm{zeeman} + H_{\textrm{scattering}} \\
  + H_{\textrm{spin-flip}} + H_{\textrm{quasispin-flip}},
\end{multline}
where $\mathcal{H}_0^\prime$ is a renormalized $\mathcal{H}_0$. Note that both $U$ and $\epsilon_f$ is renormalized to $U^\prime$ and $\epsilon_f^\prime$, and the particle-hole symmetry is broken even for $\epsilon_f = -U/2$ because of the polarized bath. 

Importantly, we see from above that there emerges an effective Zeeman splitting term $\mathcal{H}_\textrm{Zeeman} = h_z (n_{f\uparrow} - n_{f\downarrow})$, with the strength
\begin{multline}
  h_z = \sum_n\frac{|U_{n\downarrow}|^2}{E_{n,-1}-\epsilon_f} - \sum_n\frac{|U_{n\uparrow}|^2}{E_{n,0}-\epsilon_f} \\
  - \sum_n\frac{|V_{n\downarrow}|^2}{E_{n,1}+\epsilon_f+U} + \sum_n\frac{|V_{n\uparrow}|^2}{E_{n,0}+\epsilon_f+U}.
\end{multline}
The contribution of the impurity-MZM coupling to the effective Zeeman term is $h_z^{\textrm{ZM}} = -|U_{0\uparrow}|^2\left[1/\epsilon_f + 1/(\epsilon_f + U)\right]$ since $|U_{0\uparrow}|^2 = |V_{0\uparrow}|^2$, which is zero in the symmetric case $\epsilon_f = -U/2$. Fig.\ref{fig:A2} shows the strength of the Zeeman term, along with the contribution of the CdGM state
\begin{equation}
  h^{\textrm{CdGM}}_z = \frac{|U_{0\downarrow}|^2}{E_{0,-1}-\epsilon_f}
\end{equation}
It is clearly shown by Fig.S2 that  it is the CdGM state that mainly contributes to the effective Zeeman term.

The scattering term is given by
\begin{multline}
  H_{\textrm{scattering}} = \sum_{mn} W^\downarrow_{mn} n_{f\downarrow} \gamma_{m,0}^\dagger \gamma_{n,0}\\
   + \sum_{mn}\left( Z^\downarrow_{mn} n_{f\downarrow} \gamma_{m,0}^\dagger \gamma_{n,0}^\dagger + h.c. \right)\\
 + \sum_{mn}\sum_{l=\pm 1} W^\uparrow_{mn,l} n_{f\uparrow} \gamma_{m,l}^\dagger \gamma_{n,l} \\
 + \sum_{mn}\left( Z^\uparrow_{mn} n_{f\downarrow} \gamma_{m,-1}^\dagger \gamma_{n,1}^\dagger + h.c. \right)
\end{multline}
This term can be absorbed into a shift of the single-particle energy of the TSC. Besides, the quasi spin-flip term is given by
\begin{multline}
  H_{\textrm{quasispin-flip}} =\\ \sum_{mn}\left( K^{(1)}_{mn} \gamma_{m,0} \gamma_{n,1} f_\uparrow f_\downarrow + K^{(2)}_{mn} \gamma_{m,0}^\dagger \gamma_{n,-1}^\dagger f_\uparrow f_\downarrow +h.c.\right) \\
  + \sum_{mn}\left( L^{(1)}_{mn} \gamma_{m,-1}^\dagger \gamma_{n,0} f_\uparrow f_\downarrow + L^{(2)}_{mn} \gamma_{m,0}^\dagger \gamma_{n,1} f_\uparrow f_\downarrow +h.c.\right)
\end{multline}
This term is usually neglected since double occupation of the impurity site is suppressed at $U\gg 0$.

Another other interesting term is the spin flipping term $H_{\textrm{spin-flip}}$, which screens the magnetic impurity. This term is given by
\begin{multline}
  H_{\textrm{spin-flip}} =\\
   \sum_{mn}\left( J^{(1)}_{mn} \gamma_{m,0}^\dagger \gamma_{n,1} f^\dagger_\uparrow f_\downarrow + J^{(2)}_{mn} \gamma_{m,-1}^\dagger \gamma_{n,0} f^\dagger_\uparrow f_\downarrow +h.c.\right) \\
   + \sum_{mn}\left( T^{(1)}_{mn} \gamma_{m,-1}^\dagger \gamma_{n,0}^\dagger f^\dagger_\uparrow f_\downarrow + T^{(2)}_{mn} \gamma_{m,0} \gamma_{n,1} f^\dagger_\uparrow f_\downarrow +h.c.\right).
\end{multline}

\section{ \label{appendix4} Deviation from the vortex center}
When the impurity is spatially separated  from the vortex center by a distance $\mathbf{r}_0$, more OAMs are involved in the hybridization. With enlarging $\mathbf{r}_0$, one expects the impurity physics to be transformed into the vortex-free case.   We consider this question and show that the ground state can indeed be adiabatically evolved into that of the vortex-free case. We assume an isotropic coupling $V(\mathbf{r}) = V(|\mathbf{r}-\mathbf{r}_0|)$ and expand the interaction as $V(\mathbf{r}) = \sum_{m}V_{m}(r)e^{im\theta}$ ($V_{m}(r) = \frac{1}{2\pi}\int\mathrm{d}\theta\; V(\mathbf{r})e^{-im\theta}$), then the hybridization is transformed into:
\begin{equation}
\begin{split}
H_{\textrm{hyb}}&= \sum_{\sigma}\int \mathrm{d}\mathbf{r}\; V(\mathbf{r})\hat{f}_{\sigma}^\dagger\hat{c}_{\mathbf{r},\sigma}+h.c.\\
&= \sqrt{2\pi}\sum_{m n}\int \mathrm{d}r\; r V_{m}(r)\\
&\times \left\{ f_{\uparrow}^\dagger\left[ u_{n,-m\uparrow}(r)\gamma_{n,-m}+v^\ast_{n,m,\uparrow}\gamma_{n,m+\nu-1}^\dagger\right] \right.\\
&+ \left. f_{\downarrow}^\dagger \left[u_{n, -m\downarrow}(r)\gamma_{n,-m-1} + v^\ast_{nm\downarrow}\gamma_{n,m+\nu}^\dagger\right]\right\}+h.c.
\end{split}
\end{equation}
\begin{figure}
    \centering
    \includegraphics[width=0.95\linewidth]{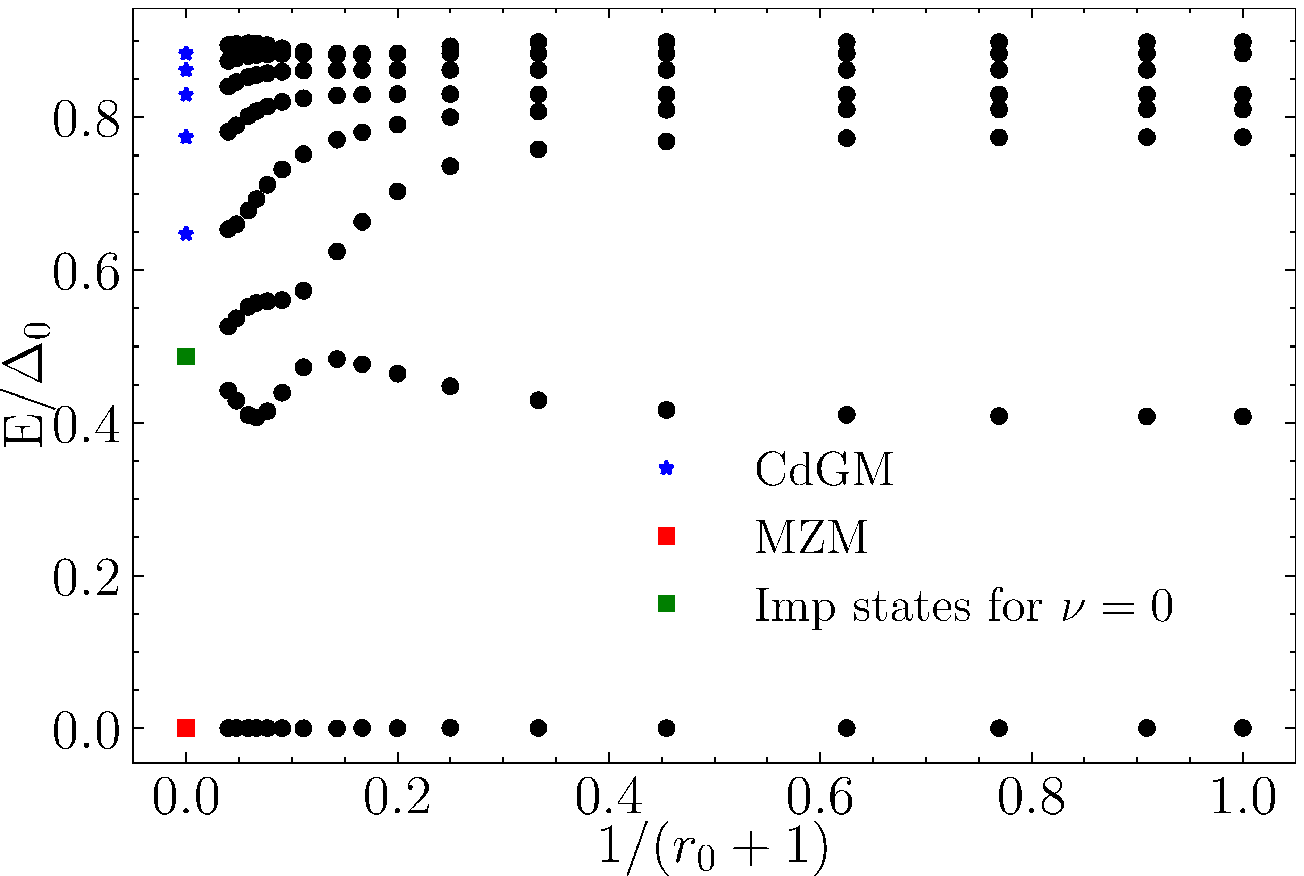}
    \caption{Single particle in-gap spectrum as a function of $1/(r_0+1)$, where $r_0$ is the distance between the impurity and the vortex center. At $r_0\to \infty$, the impurity state is obtained from $\nu = 0$, and the MZM and CdGM states are obtained from the vortex state without impurity.}
    \label{fig:A3}
\end{figure}
This is essentially a two-dimensional model, which is generally difficult to solve except for $U=0$. For $U=0$, the single particle spectrum is show in Fig.\ref{fig:A3} as a function of the distance between the impurity and the vortex center. Here, the truncation of the maximum OAM is chosen as $m_{\textrm{max}} = 170$ for $r_0=24$. The data for $r_0\to \infty$ is obtained from vortex free case and vortex state without impurity, respectively. As we observe in Fig.S4, when the impurity is gradually separated from the vortex center, the energies of the two polarized in-gap impurity states evolve continuously and non-monotonically. Meanwhile, they become closer and closer to the degenerate doublet in the vortex-free case. The fluctuating behavior of the energy in Fig.S3 is mainly caused by the spatial distribution of the CdGM and MZM states.

\section{Symmetry analysis accounting for the RS-LM crossover}
\begin{figure}[b]
  \includegraphics[width=0.95\linewidth]{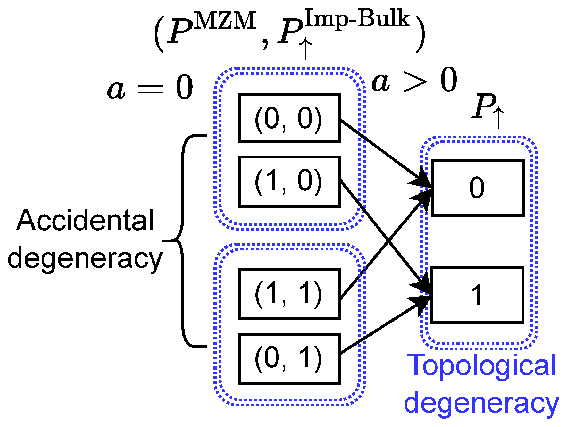}
  \caption{Schematic plot to show that at QPT the accident degeneracy is lifted by the impurity-MZM coupling between $P^\textrm{Bulk+Imp}_\uparrow = 0$ and $1$ sectors, leaving only the topological degeneracy.}
  \label{fig:A4}
\end{figure}

In the main text we concluded that the RS-LM crossover is a general result because of the symmetry at the critical point. We demonstrate this point more clearly in this section. We intentionally introduced a controlling factor $a$ in front of the  impurity-MZM coupling to keep track of its effect on the transition point. In the non-coupling limit ($a=0$), a QPT occurs at $U\approx 0.15$ with the other parameters given in the main text. An infinitesimal $a$ is sufficient  to turn the QPT into a crossover, manifesting as an avoided level crossing, as we showed in the manuscript.
 
Here we present a symmetry analysis to account for the avoided level crossing. For $a = 0$ at the QPT, the ground state has quadruple degeneracy. $N_\downarrow$ is the same for those four states, so it is irrelevant in the following analysis. Taking the free MZM into account, the parity in the bulk-impurity system ($P^\textrm{Imp-Bulk}_\uparrow$) and the parity of the MZM ($P^\textrm{MZM}$) are conserved separately. The total parity is then given by $P^\uparrow = (P^\textrm{Imp-Bulk}_\uparrow + P^\textrm{MZM})$(\textrm{mod} 2). The four degenerated states can be separated into two sectors as shown in Fig.\ref{fig:A4}. One is $(P^\textrm{MZM}, P^\textrm{Imp-Bulk}_\uparrow) = \left\{(0, 0), (1, 0)\right\}$, and the other is $(P^\textrm{MZM}, P^\textrm{Imp-Bulk}_\uparrow) =\left\{ (0, 1), (1, 1)\right\}$. Within each sector, the remaining degeneracy is of topological origin, as a result of the presence of MZM. However, the degeneracy of the states in between the two sector is not topological but an accidental one. We find that the impurity-MZM coupling off-diagonal entries  that mix the $P^\textrm{Bulk+Imp}_\uparrow = 0$ and $1$ sectors, which in turn lifts the accident degeneracy, leaving only the topological degeneracy. This results in the avoided level crossing numerically observed in the main text.
\bibliography{reference}

\begin{thebibliography}{38}%
\makeatletter
\providecommand \@ifxundefined [1]{%
 \@ifx{#1\undefined}
}%
\providecommand \@ifnum [1]{%
 \ifnum #1\expandafter \@firstoftwo
 \else \expandafter \@secondoftwo
 \fi
}%
\providecommand \@ifx [1]{%
 \ifx #1\expandafter \@firstoftwo
 \else \expandafter \@secondoftwo
 \fi
}%
\providecommand \natexlab [1]{#1}%
\providecommand \enquote  [1]{``#1''}%
\providecommand \bibnamefont  [1]{#1}%
\providecommand \bibfnamefont [1]{#1}%
\providecommand \citenamefont [1]{#1}%
\providecommand \href@noop [0]{\@secondoftwo}%
\providecommand \href [0]{\begingroup \@sanitize@url \@href}%
\providecommand \@href[1]{\@@startlink{#1}\@@href}%
\providecommand \@@href[1]{\endgroup#1\@@endlink}%
\providecommand \@sanitize@url [0]{\catcode `\\12\catcode `\$12\catcode
  `\&12\catcode `\#12\catcode `\^12\catcode `\_12\catcode `\%12\relax}%
\providecommand \@@startlink[1]{}%
\providecommand \@@endlink[0]{}%
\providecommand \url  [0]{\begingroup\@sanitize@url \@url }%
\providecommand \@url [1]{\endgroup\@href {#1}{\urlprefix }}%
\providecommand \urlprefix  [0]{URL }%
\providecommand \Eprint [0]{\href }%
\providecommand \doibase [0]{https://doi.org/}%
\providecommand \selectlanguage [0]{\@gobble}%
\providecommand \bibinfo  [0]{\@secondoftwo}%
\providecommand \bibfield  [0]{\@secondoftwo}%
\providecommand \translation [1]{[#1]}%
\providecommand \BibitemOpen [0]{}%
\providecommand \bibitemStop [0]{}%
\providecommand \bibitemNoStop [0]{.\EOS\space}%
\providecommand \EOS [0]{\spacefactor3000\relax}%
\providecommand \BibitemShut  [1]{\csname bibitem#1\endcsname}%
\let\auto@bib@innerbib\@empty
\bibitem [{\citenamefont {Elliott}\ and\ \citenamefont
  {Franz}(2015)}]{RevModPhys.87.137}%
  \BibitemOpen
  \bibfield  {author} {\bibinfo {author} {\bibfnamefont {S.~R.}\ \bibnamefont
  {Elliott}}\ and\ \bibinfo {author} {\bibfnamefont {M.}~\bibnamefont
  {Franz}},\ }\bibfield  {title} {\bibinfo {title} {Colloquium: Majorana
  fermions in nuclear, particle, and solid-state physics},\ }\href
  {https://doi.org/10.1103/RevModPhys.87.137} {\bibfield  {journal} {\bibinfo
  {journal} {Rev. Mod. Phys.}\ }\textbf {\bibinfo {volume} {87}},\ \bibinfo
  {pages} {137} (\bibinfo {year} {2015})}\BibitemShut {NoStop}%
\bibitem [{\citenamefont {Nayak}\ \emph {et~al.}(2008)\citenamefont {Nayak},
  \citenamefont {Simon}, \citenamefont {Stern}, \citenamefont {Freedman},\ and\
  \citenamefont {Das~Sarma}}]{RevModPhys.80.1083}%
  \BibitemOpen
  \bibfield  {author} {\bibinfo {author} {\bibfnamefont {C.}~\bibnamefont
  {Nayak}}, \bibinfo {author} {\bibfnamefont {S.~H.}\ \bibnamefont {Simon}},
  \bibinfo {author} {\bibfnamefont {A.}~\bibnamefont {Stern}}, \bibinfo
  {author} {\bibfnamefont {M.}~\bibnamefont {Freedman}},\ and\ \bibinfo
  {author} {\bibfnamefont {S.}~\bibnamefont {Das~Sarma}},\ }\bibfield  {title}
  {\bibinfo {title} {Non-abelian anyons and topological quantum computation},\
  }\href {https://doi.org/10.1103/RevModPhys.80.1083} {\bibfield  {journal}
  {\bibinfo  {journal} {Rev. Mod. Phys.}\ }\textbf {\bibinfo {volume} {80}},\
  \bibinfo {pages} {1083} (\bibinfo {year} {2008})}\BibitemShut {NoStop}%
\bibitem [{\citenamefont {Zhang}\ \emph {et~al.}(2021)\citenamefont {Zhang},
  \citenamefont {Bao}, \citenamefont {Chen}, \citenamefont {Li}, \citenamefont
  {Lu}, \citenamefont {Hu}, \citenamefont {Yang}, \citenamefont {Zhao},
  \citenamefont {Yan}, \citenamefont {Dong}, \citenamefont {Wang},
  \citenamefont {Zhang},\ and\ \citenamefont {Feng}}]{PhysRevLett.126.127001}%
  \BibitemOpen
  \bibfield  {author} {\bibinfo {author} {\bibfnamefont {T.}~\bibnamefont
  {Zhang}}, \bibinfo {author} {\bibfnamefont {W.}~\bibnamefont {Bao}}, \bibinfo
  {author} {\bibfnamefont {C.}~\bibnamefont {Chen}}, \bibinfo {author}
  {\bibfnamefont {D.}~\bibnamefont {Li}}, \bibinfo {author} {\bibfnamefont
  {Z.}~\bibnamefont {Lu}}, \bibinfo {author} {\bibfnamefont {Y.}~\bibnamefont
  {Hu}}, \bibinfo {author} {\bibfnamefont {W.}~\bibnamefont {Yang}}, \bibinfo
  {author} {\bibfnamefont {D.}~\bibnamefont {Zhao}}, \bibinfo {author}
  {\bibfnamefont {Y.}~\bibnamefont {Yan}}, \bibinfo {author} {\bibfnamefont
  {X.}~\bibnamefont {Dong}}, \bibinfo {author} {\bibfnamefont {Q.-H.}\
  \bibnamefont {Wang}}, \bibinfo {author} {\bibfnamefont {T.}~\bibnamefont
  {Zhang}},\ and\ \bibinfo {author} {\bibfnamefont {D.}~\bibnamefont {Feng}},\
  }\bibfield  {title} {\bibinfo {title} {Observation of distinct spatial
  distributions of the zero and nonzero energy vortex modes in
  $({\mathrm{li}}_{0.84}{\mathrm{fe}}_{0.16})\mathrm{OHFeSe}$},\ }\href
  {https://doi.org/10.1103/PhysRevLett.126.127001} {\bibfield  {journal}
  {\bibinfo  {journal} {Phys. Rev. Lett.}\ }\textbf {\bibinfo {volume} {126}},\
  \bibinfo {pages} {127001} (\bibinfo {year} {2021})}\BibitemShut {NoStop}%
\bibitem [{\citenamefont {Sun}\ \emph {et~al.}(2016)\citenamefont {Sun},
  \citenamefont {Zhang}, \citenamefont {Hu}, \citenamefont {Li}, \citenamefont
  {Wang}, \citenamefont {Ma}, \citenamefont {Xu}, \citenamefont {Gao},
  \citenamefont {Guan}, \citenamefont {Li}, \citenamefont {Liu}, \citenamefont
  {Qian}, \citenamefont {Zhou}, \citenamefont {Fu}, \citenamefont {Li},
  \citenamefont {Zhang},\ and\ \citenamefont {Jia}}]{PhysRevLett.116.257003}%
  \BibitemOpen
  \bibfield  {author} {\bibinfo {author} {\bibfnamefont {H.-H.}\ \bibnamefont
  {Sun}}, \bibinfo {author} {\bibfnamefont {K.-W.}\ \bibnamefont {Zhang}},
  \bibinfo {author} {\bibfnamefont {L.-H.}\ \bibnamefont {Hu}}, \bibinfo
  {author} {\bibfnamefont {C.}~\bibnamefont {Li}}, \bibinfo {author}
  {\bibfnamefont {G.-Y.}\ \bibnamefont {Wang}}, \bibinfo {author}
  {\bibfnamefont {H.-Y.}\ \bibnamefont {Ma}}, \bibinfo {author} {\bibfnamefont
  {Z.-A.}\ \bibnamefont {Xu}}, \bibinfo {author} {\bibfnamefont {C.-L.}\
  \bibnamefont {Gao}}, \bibinfo {author} {\bibfnamefont {D.-D.}\ \bibnamefont
  {Guan}}, \bibinfo {author} {\bibfnamefont {Y.-Y.}\ \bibnamefont {Li}},
  \bibinfo {author} {\bibfnamefont {C.}~\bibnamefont {Liu}}, \bibinfo {author}
  {\bibfnamefont {D.}~\bibnamefont {Qian}}, \bibinfo {author} {\bibfnamefont
  {Y.}~\bibnamefont {Zhou}}, \bibinfo {author} {\bibfnamefont {L.}~\bibnamefont
  {Fu}}, \bibinfo {author} {\bibfnamefont {S.-C.}\ \bibnamefont {Li}}, \bibinfo
  {author} {\bibfnamefont {F.-C.}\ \bibnamefont {Zhang}},\ and\ \bibinfo
  {author} {\bibfnamefont {J.-F.}\ \bibnamefont {Jia}},\ }\bibfield  {title}
  {\bibinfo {title} {Majorana zero mode detected with spin selective andreev
  reflection in the vortex of a topological superconductor},\ }\href
  {https://doi.org/10.1103/PhysRevLett.116.257003} {\bibfield  {journal}
  {\bibinfo  {journal} {Phys. Rev. Lett.}\ }\textbf {\bibinfo {volume} {116}},\
  \bibinfo {pages} {257003} (\bibinfo {year} {2016})}\BibitemShut {NoStop}%
\bibitem [{\citenamefont {Xu}\ \emph {et~al.}(2016)\citenamefont {Xu},
  \citenamefont {Lian}, \citenamefont {Tang}, \citenamefont {Qi},\ and\
  \citenamefont {Zhang}}]{PhysRevLett.117.047001}%
  \BibitemOpen
  \bibfield  {author} {\bibinfo {author} {\bibfnamefont {G.}~\bibnamefont
  {Xu}}, \bibinfo {author} {\bibfnamefont {B.}~\bibnamefont {Lian}}, \bibinfo
  {author} {\bibfnamefont {P.}~\bibnamefont {Tang}}, \bibinfo {author}
  {\bibfnamefont {X.-L.}\ \bibnamefont {Qi}},\ and\ \bibinfo {author}
  {\bibfnamefont {S.-C.}\ \bibnamefont {Zhang}},\ }\bibfield  {title} {\bibinfo
  {title} {Topological superconductivity on the surface of fe-based
  superconductors},\ }\href {https://doi.org/10.1103/PhysRevLett.117.047001}
  {\bibfield  {journal} {\bibinfo  {journal} {Phys. Rev. Lett.}\ }\textbf
  {\bibinfo {volume} {117}},\ \bibinfo {pages} {047001} (\bibinfo {year}
  {2016})}\BibitemShut {NoStop}%
\bibitem [{\citenamefont {Liu}\ \emph {et~al.}(2018)\citenamefont {Liu},
  \citenamefont {Chen}, \citenamefont {Zhang}, \citenamefont {Peng},
  \citenamefont {Yan}, \citenamefont {Wen}, \citenamefont {Lou}, \citenamefont
  {Huang}, \citenamefont {Tian}, \citenamefont {Dong}, \citenamefont {Wang},
  \citenamefont {Bao}, \citenamefont {Wang}, \citenamefont {Yin}, \citenamefont
  {Zhao},\ and\ \citenamefont {Feng}}]{PhysRevX.8.041056}%
  \BibitemOpen
  \bibfield  {author} {\bibinfo {author} {\bibfnamefont {Q.}~\bibnamefont
  {Liu}}, \bibinfo {author} {\bibfnamefont {C.}~\bibnamefont {Chen}}, \bibinfo
  {author} {\bibfnamefont {T.}~\bibnamefont {Zhang}}, \bibinfo {author}
  {\bibfnamefont {R.}~\bibnamefont {Peng}}, \bibinfo {author} {\bibfnamefont
  {Y.-J.}\ \bibnamefont {Yan}}, \bibinfo {author} {\bibfnamefont {C.-H.-P.}\
  \bibnamefont {Wen}}, \bibinfo {author} {\bibfnamefont {X.}~\bibnamefont
  {Lou}}, \bibinfo {author} {\bibfnamefont {Y.-L.}\ \bibnamefont {Huang}},
  \bibinfo {author} {\bibfnamefont {J.-P.}\ \bibnamefont {Tian}}, \bibinfo
  {author} {\bibfnamefont {X.-L.}\ \bibnamefont {Dong}}, \bibinfo {author}
  {\bibfnamefont {G.-W.}\ \bibnamefont {Wang}}, \bibinfo {author}
  {\bibfnamefont {W.-C.}\ \bibnamefont {Bao}}, \bibinfo {author} {\bibfnamefont
  {Q.-H.}\ \bibnamefont {Wang}}, \bibinfo {author} {\bibfnamefont {Z.-P.}\
  \bibnamefont {Yin}}, \bibinfo {author} {\bibfnamefont {Z.-X.}\ \bibnamefont
  {Zhao}},\ and\ \bibinfo {author} {\bibfnamefont {D.-L.}\ \bibnamefont
  {Feng}},\ }\bibfield  {title} {\bibinfo {title} {Robust and clean majorana
  zero mode in the vortex core of high-temperature superconductor
  $\mathbf{(}{\mathrm{li}}_{0.84}{\mathrm{fe}}_{0.16}\mathbf{)}\mathrm{OHFeSe}$},\
  }\href {https://doi.org/10.1103/PhysRevX.8.041056} {\bibfield  {journal}
  {\bibinfo  {journal} {Phys. Rev. X}\ }\textbf {\bibinfo {volume} {8}},\
  \bibinfo {pages} {041056} (\bibinfo {year} {2018})}\BibitemShut {NoStop}%
\bibitem [{\citenamefont {Wang}\ \emph {et~al.}(2018)\citenamefont {Wang},
  \citenamefont {Kong}, \citenamefont {Fan}, \citenamefont {Chen},
  \citenamefont {Zhu}, \citenamefont {Liu}, \citenamefont {Cao}, \citenamefont
  {Sun}, \citenamefont {Du}, \citenamefont {Schneeloch} \emph
  {et~al.}}]{wang2018evidence}%
  \BibitemOpen
  \bibfield  {author} {\bibinfo {author} {\bibfnamefont {D.}~\bibnamefont
  {Wang}}, \bibinfo {author} {\bibfnamefont {L.}~\bibnamefont {Kong}}, \bibinfo
  {author} {\bibfnamefont {P.}~\bibnamefont {Fan}}, \bibinfo {author}
  {\bibfnamefont {H.}~\bibnamefont {Chen}}, \bibinfo {author} {\bibfnamefont
  {S.}~\bibnamefont {Zhu}}, \bibinfo {author} {\bibfnamefont {W.}~\bibnamefont
  {Liu}}, \bibinfo {author} {\bibfnamefont {L.}~\bibnamefont {Cao}}, \bibinfo
  {author} {\bibfnamefont {Y.}~\bibnamefont {Sun}}, \bibinfo {author}
  {\bibfnamefont {S.}~\bibnamefont {Du}}, \bibinfo {author} {\bibfnamefont
  {J.}~\bibnamefont {Schneeloch}}, \emph {et~al.},\ }\bibfield  {title}
  {\bibinfo {title} {Evidence for majorana bound states in an iron-based
  superconductor},\ }\href@noop {} {\bibfield  {journal} {\bibinfo  {journal}
  {Science}\ }\textbf {\bibinfo {volume} {362}},\ \bibinfo {pages} {333}
  (\bibinfo {year} {2018})}\BibitemShut {NoStop}%
\bibitem [{\citenamefont {Zhang}\ \emph {et~al.}(2018)\citenamefont {Zhang},
  \citenamefont {Yaji}, \citenamefont {Hashimoto}, \citenamefont {Ota},
  \citenamefont {Kondo}, \citenamefont {Okazaki}, \citenamefont {Wang},
  \citenamefont {Wen}, \citenamefont {Gu}, \citenamefont {Ding} \emph
  {et~al.}}]{zhang2018observation}%
  \BibitemOpen
  \bibfield  {author} {\bibinfo {author} {\bibfnamefont {P.}~\bibnamefont
  {Zhang}}, \bibinfo {author} {\bibfnamefont {K.}~\bibnamefont {Yaji}},
  \bibinfo {author} {\bibfnamefont {T.}~\bibnamefont {Hashimoto}}, \bibinfo
  {author} {\bibfnamefont {Y.}~\bibnamefont {Ota}}, \bibinfo {author}
  {\bibfnamefont {T.}~\bibnamefont {Kondo}}, \bibinfo {author} {\bibfnamefont
  {K.}~\bibnamefont {Okazaki}}, \bibinfo {author} {\bibfnamefont
  {Z.}~\bibnamefont {Wang}}, \bibinfo {author} {\bibfnamefont {J.}~\bibnamefont
  {Wen}}, \bibinfo {author} {\bibfnamefont {G.}~\bibnamefont {Gu}}, \bibinfo
  {author} {\bibfnamefont {H.}~\bibnamefont {Ding}}, \emph {et~al.},\
  }\bibfield  {title} {\bibinfo {title} {Observation of topological
  superconductivity on the surface of an iron-based superconductor},\
  }\href@noop {} {\bibfield  {journal} {\bibinfo  {journal} {Science}\ }\textbf
  {\bibinfo {volume} {360}},\ \bibinfo {pages} {182} (\bibinfo {year}
  {2018})}\BibitemShut {NoStop}%
\bibitem [{\citenamefont {Yu}(1965)}]{yu1965bound}%
  \BibitemOpen
  \bibfield  {author} {\bibinfo {author} {\bibfnamefont {L.}~\bibnamefont
  {Yu}},\ }\bibfield  {title} {\bibinfo {title} {Bound state in superconductors
  with paramagnetic impurities},\ }\href@noop {} {\bibfield  {journal}
  {\bibinfo  {journal} {Acta Phys. Sin}\ }\textbf {\bibinfo {volume} {21}},\
  \bibinfo {pages} {75} (\bibinfo {year} {1965})}\BibitemShut {NoStop}%
\bibitem [{\citenamefont {Shiba}(1968)}]{shiba1968classical}%
  \BibitemOpen
  \bibfield  {author} {\bibinfo {author} {\bibfnamefont {H.}~\bibnamefont
  {Shiba}},\ }\bibfield  {title} {\bibinfo {title} {Classical spins in
  superconductors},\ }\href@noop {} {\bibfield  {journal} {\bibinfo  {journal}
  {Progress of theoretical Physics}\ }\textbf {\bibinfo {volume} {40}},\
  \bibinfo {pages} {435} (\bibinfo {year} {1968})}\BibitemShut {NoStop}%
\bibitem [{\citenamefont {Wang}\ \emph {et~al.}(2019)\citenamefont {Wang},
  \citenamefont {Su}, \citenamefont {Zhu}, \citenamefont {Ting}, \citenamefont
  {Li}, \citenamefont {Chen}, \citenamefont {Wang},\ and\ \citenamefont
  {Wang}}]{PhysRevLett.122.087001}%
  \BibitemOpen
  \bibfield  {author} {\bibinfo {author} {\bibfnamefont {R.}~\bibnamefont
  {Wang}}, \bibinfo {author} {\bibfnamefont {W.}~\bibnamefont {Su}}, \bibinfo
  {author} {\bibfnamefont {J.-X.}\ \bibnamefont {Zhu}}, \bibinfo {author}
  {\bibfnamefont {C.~S.}\ \bibnamefont {Ting}}, \bibinfo {author}
  {\bibfnamefont {H.}~\bibnamefont {Li}}, \bibinfo {author} {\bibfnamefont
  {C.}~\bibnamefont {Chen}}, \bibinfo {author} {\bibfnamefont {B.}~\bibnamefont
  {Wang}},\ and\ \bibinfo {author} {\bibfnamefont {X.}~\bibnamefont {Wang}},\
  }\bibfield  {title} {\bibinfo {title} {Kondo signatures of a quantum magnetic
  impurity in topological superconductors},\ }\href
  {https://doi.org/10.1103/PhysRevLett.122.087001} {\bibfield  {journal}
  {\bibinfo  {journal} {Phys. Rev. Lett.}\ }\textbf {\bibinfo {volume} {122}},\
  \bibinfo {pages} {087001} (\bibinfo {year} {2019})}\BibitemShut {NoStop}%
\bibitem [{\citenamefont {Gao}\ \emph {et~al.}(2017)\citenamefont {Gao},
  \citenamefont {Yu}, \citenamefont {Zhou}, \citenamefont {Huang},\ and\
  \citenamefont {Wang}}]{PhysRevB.96.220507}%
  \BibitemOpen
  \bibfield  {author} {\bibinfo {author} {\bibfnamefont {Y.}~\bibnamefont
  {Gao}}, \bibinfo {author} {\bibfnamefont {Y.}~\bibnamefont {Yu}}, \bibinfo
  {author} {\bibfnamefont {T.}~\bibnamefont {Zhou}}, \bibinfo {author}
  {\bibfnamefont {H.}~\bibnamefont {Huang}},\ and\ \bibinfo {author}
  {\bibfnamefont {Q.-H.}\ \bibnamefont {Wang}},\ }\bibfield  {title} {\bibinfo
  {title} {In-gap bound states induced by a single nonmagnetic impurity in
  sign-preserving $s$-wave superconductors with incipient bands},\ }\href
  {https://doi.org/10.1103/PhysRevB.96.220507} {\bibfield  {journal} {\bibinfo
  {journal} {Phys. Rev. B}\ }\textbf {\bibinfo {volume} {96}},\ \bibinfo
  {pages} {220507} (\bibinfo {year} {2017})}\BibitemShut {NoStop}%
\bibitem [{\citenamefont {Cornils}\ \emph {et~al.}(2017)\citenamefont
  {Cornils}, \citenamefont {Kamlapure}, \citenamefont {Zhou}, \citenamefont
  {Pradhan}, \citenamefont {Khajetoorians}, \citenamefont {Fransson},
  \citenamefont {Wiebe},\ and\ \citenamefont
  {Wiesendanger}}]{PhysRevLett.119.197002}%
  \BibitemOpen
  \bibfield  {author} {\bibinfo {author} {\bibfnamefont {L.}~\bibnamefont
  {Cornils}}, \bibinfo {author} {\bibfnamefont {A.}~\bibnamefont {Kamlapure}},
  \bibinfo {author} {\bibfnamefont {L.}~\bibnamefont {Zhou}}, \bibinfo {author}
  {\bibfnamefont {S.}~\bibnamefont {Pradhan}}, \bibinfo {author} {\bibfnamefont
  {A.~A.}\ \bibnamefont {Khajetoorians}}, \bibinfo {author} {\bibfnamefont
  {J.}~\bibnamefont {Fransson}}, \bibinfo {author} {\bibfnamefont
  {J.}~\bibnamefont {Wiebe}},\ and\ \bibinfo {author} {\bibfnamefont
  {R.}~\bibnamefont {Wiesendanger}},\ }\bibfield  {title} {\bibinfo {title}
  {Spin-resolved spectroscopy of the yu-shiba-rusinov states of individual
  atoms},\ }\href {https://doi.org/10.1103/PhysRevLett.119.197002} {\bibfield
  {journal} {\bibinfo  {journal} {Phys. Rev. Lett.}\ }\textbf {\bibinfo
  {volume} {119}},\ \bibinfo {pages} {197002} (\bibinfo {year}
  {2017})}\BibitemShut {NoStop}%
\bibitem [{\citenamefont {Cort{\'e}s-del R{\'\i}o}\ \emph
  {et~al.}(2021)\citenamefont {Cort{\'e}s-del R{\'\i}o}, \citenamefont {Lado},
  \citenamefont {Cherkez}, \citenamefont {Mallet}, \citenamefont {Veuillen},
  \citenamefont {Cuevas}, \citenamefont {G{\'o}mez-Rodr{\'\i}guez},
  \citenamefont {Fern{\'a}ndez-Rossier},\ and\ \citenamefont
  {Brihuega}}]{cortes2021observation}%
  \BibitemOpen
  \bibfield  {author} {\bibinfo {author} {\bibfnamefont {E.}~\bibnamefont
  {Cort{\'e}s-del R{\'\i}o}}, \bibinfo {author} {\bibfnamefont {J.~L.}\
  \bibnamefont {Lado}}, \bibinfo {author} {\bibfnamefont {V.}~\bibnamefont
  {Cherkez}}, \bibinfo {author} {\bibfnamefont {P.}~\bibnamefont {Mallet}},
  \bibinfo {author} {\bibfnamefont {J.-Y.}\ \bibnamefont {Veuillen}}, \bibinfo
  {author} {\bibfnamefont {J.~C.}\ \bibnamefont {Cuevas}}, \bibinfo {author}
  {\bibfnamefont {J.~M.}\ \bibnamefont {G{\'o}mez-Rodr{\'\i}guez}}, \bibinfo
  {author} {\bibfnamefont {J.}~\bibnamefont {Fern{\'a}ndez-Rossier}},\ and\
  \bibinfo {author} {\bibfnamefont {I.}~\bibnamefont {Brihuega}},\ }\bibfield
  {title} {\bibinfo {title} {Observation of yu--shiba--rusinov states in
  superconducting graphene},\ }\href@noop {} {\bibfield  {journal} {\bibinfo
  {journal} {Advanced Materials}\ }\textbf {\bibinfo {volume} {33}},\ \bibinfo
  {pages} {2008113} (\bibinfo {year} {2021})}\BibitemShut {NoStop}%
\bibitem [{\citenamefont {Jiang}\ \emph {et~al.}(2019)\citenamefont {Jiang},
  \citenamefont {Dai},\ and\ \citenamefont {Wang}}]{PhysRevX.9.011033}%
  \BibitemOpen
  \bibfield  {author} {\bibinfo {author} {\bibfnamefont {K.}~\bibnamefont
  {Jiang}}, \bibinfo {author} {\bibfnamefont {X.}~\bibnamefont {Dai}},\ and\
  \bibinfo {author} {\bibfnamefont {Z.}~\bibnamefont {Wang}},\ }\bibfield
  {title} {\bibinfo {title} {Quantum anomalous vortex and majorana zero mode in
  iron-based superconductor fe(te,se)},\ }\href
  {https://doi.org/10.1103/PhysRevX.9.011033} {\bibfield  {journal} {\bibinfo
  {journal} {Phys. Rev. X}\ }\textbf {\bibinfo {volume} {9}},\ \bibinfo {pages}
  {011033} (\bibinfo {year} {2019})}\BibitemShut {NoStop}%
\bibitem [{\citenamefont {Qin}\ \emph {et~al.}(2020)\citenamefont {Qin},
  \citenamefont {Fang}, \citenamefont {Zhang},\ and\ \citenamefont
  {Hu}}]{qin2020topological}%
  \BibitemOpen
  \bibfield  {author} {\bibinfo {author} {\bibfnamefont {S.}~\bibnamefont
  {Qin}}, \bibinfo {author} {\bibfnamefont {C.}~\bibnamefont {Fang}}, \bibinfo
  {author} {\bibfnamefont {F.-c.}\ \bibnamefont {Zhang}},\ and\ \bibinfo
  {author} {\bibfnamefont {J.}~\bibnamefont {Hu}},\ }\bibfield  {title}
  {\bibinfo {title} {Topological phase transitions of superconducting vortex
  bound states driven by impurities},\ }\href@noop {} {\bibfield  {journal}
  {\bibinfo  {journal} {arXiv preprint arXiv:2010.09382}\ } (\bibinfo {year}
  {2020})}\BibitemShut {NoStop}%
\bibitem [{\citenamefont {Yin}\ \emph {et~al.}(2015)\citenamefont {Yin},
  \citenamefont {Wu}, \citenamefont {Wang}, \citenamefont {Ye}, \citenamefont
  {Gong}, \citenamefont {Hou}, \citenamefont {Shan}, \citenamefont {Li},
  \citenamefont {Liang}, \citenamefont {Wu} \emph
  {et~al.}}]{yin2015observation}%
  \BibitemOpen
  \bibfield  {author} {\bibinfo {author} {\bibfnamefont {J.}~\bibnamefont
  {Yin}}, \bibinfo {author} {\bibfnamefont {Z.}~\bibnamefont {Wu}}, \bibinfo
  {author} {\bibfnamefont {J.}~\bibnamefont {Wang}}, \bibinfo {author}
  {\bibfnamefont {Z.}~\bibnamefont {Ye}}, \bibinfo {author} {\bibfnamefont
  {J.}~\bibnamefont {Gong}}, \bibinfo {author} {\bibfnamefont {X.}~\bibnamefont
  {Hou}}, \bibinfo {author} {\bibfnamefont {L.}~\bibnamefont {Shan}}, \bibinfo
  {author} {\bibfnamefont {A.}~\bibnamefont {Li}}, \bibinfo {author}
  {\bibfnamefont {X.}~\bibnamefont {Liang}}, \bibinfo {author} {\bibfnamefont
  {X.}~\bibnamefont {Wu}}, \emph {et~al.},\ }\bibfield  {title} {\bibinfo
  {title} {Observation of a robust zero-energy bound state in iron-based
  superconductor fe (te, se)},\ }\href@noop {} {\bibfield  {journal} {\bibinfo
  {journal} {Nature Physics}\ }\textbf {\bibinfo {volume} {11}},\ \bibinfo
  {pages} {543} (\bibinfo {year} {2015})}\BibitemShut {NoStop}%
\bibitem [{\citenamefont {Fan}\ \emph {et~al.}(2021)\citenamefont {Fan},
  \citenamefont {Yang}, \citenamefont {Qian}, \citenamefont {Chen},
  \citenamefont {Zhang}, \citenamefont {Li}, \citenamefont {Huang},
  \citenamefont {Xing}, \citenamefont {Kong}, \citenamefont {Liu} \emph
  {et~al.}}]{fan2021observation}%
  \BibitemOpen
  \bibfield  {author} {\bibinfo {author} {\bibfnamefont {P.}~\bibnamefont
  {Fan}}, \bibinfo {author} {\bibfnamefont {F.}~\bibnamefont {Yang}}, \bibinfo
  {author} {\bibfnamefont {G.}~\bibnamefont {Qian}}, \bibinfo {author}
  {\bibfnamefont {H.}~\bibnamefont {Chen}}, \bibinfo {author} {\bibfnamefont
  {Y.-Y.}\ \bibnamefont {Zhang}}, \bibinfo {author} {\bibfnamefont
  {G.}~\bibnamefont {Li}}, \bibinfo {author} {\bibfnamefont {Z.}~\bibnamefont
  {Huang}}, \bibinfo {author} {\bibfnamefont {Y.}~\bibnamefont {Xing}},
  \bibinfo {author} {\bibfnamefont {L.}~\bibnamefont {Kong}}, \bibinfo {author}
  {\bibfnamefont {W.}~\bibnamefont {Liu}}, \emph {et~al.},\ }\bibfield  {title}
  {\bibinfo {title} {Observation of magnetic adatom-induced majorana vortex and
  its hybridization with field-induced majorana vortex in an iron-based
  superconductor},\ }\href@noop {} {\bibfield  {journal} {\bibinfo  {journal}
  {Nature communications}\ }\textbf {\bibinfo {volume} {12}},\ \bibinfo {pages}
  {1} (\bibinfo {year} {2021})}\BibitemShut {NoStop}%
\bibitem [{\citenamefont {Kong}\ \emph {et~al.}(2021)\citenamefont {Kong},
  \citenamefont {Cao}, \citenamefont {Zhu}, \citenamefont {Papaj},
  \citenamefont {Dai}, \citenamefont {Li}, \citenamefont {Fan}, \citenamefont
  {Liu}, \citenamefont {Yang}, \citenamefont {Wang} \emph
  {et~al.}}]{kong2021majorana}%
  \BibitemOpen
  \bibfield  {author} {\bibinfo {author} {\bibfnamefont {L.}~\bibnamefont
  {Kong}}, \bibinfo {author} {\bibfnamefont {L.}~\bibnamefont {Cao}}, \bibinfo
  {author} {\bibfnamefont {S.}~\bibnamefont {Zhu}}, \bibinfo {author}
  {\bibfnamefont {M.}~\bibnamefont {Papaj}}, \bibinfo {author} {\bibfnamefont
  {G.}~\bibnamefont {Dai}}, \bibinfo {author} {\bibfnamefont {G.}~\bibnamefont
  {Li}}, \bibinfo {author} {\bibfnamefont {P.}~\bibnamefont {Fan}}, \bibinfo
  {author} {\bibfnamefont {W.}~\bibnamefont {Liu}}, \bibinfo {author}
  {\bibfnamefont {F.}~\bibnamefont {Yang}}, \bibinfo {author} {\bibfnamefont
  {X.}~\bibnamefont {Wang}}, \emph {et~al.},\ }\bibfield  {title} {\bibinfo
  {title} {Majorana zero modes in impurity-assisted vortex of lifeas
  superconductor},\ }\href@noop {} {\bibfield  {journal} {\bibinfo  {journal}
  {Nature Communications}\ }\textbf {\bibinfo {volume} {12}},\ \bibinfo {pages}
  {1} (\bibinfo {year} {2021})}\BibitemShut {NoStop}%
\bibitem [{\citenamefont {Wang}\ \emph {et~al.}(2021)\citenamefont {Wang},
  \citenamefont {Wiebe}, \citenamefont {Zhong}, \citenamefont {Gu},\ and\
  \citenamefont {Wiesendanger}}]{PhysRevLett.126.076802}%
  \BibitemOpen
  \bibfield  {author} {\bibinfo {author} {\bibfnamefont {D.}~\bibnamefont
  {Wang}}, \bibinfo {author} {\bibfnamefont {J.}~\bibnamefont {Wiebe}},
  \bibinfo {author} {\bibfnamefont {R.}~\bibnamefont {Zhong}}, \bibinfo
  {author} {\bibfnamefont {G.}~\bibnamefont {Gu}},\ and\ \bibinfo {author}
  {\bibfnamefont {R.}~\bibnamefont {Wiesendanger}},\ }\bibfield  {title}
  {\bibinfo {title} {Spin-polarized yu-shiba-rusinov states in an iron-based
  superconductor},\ }\href {https://doi.org/10.1103/PhysRevLett.126.076802}
  {\bibfield  {journal} {\bibinfo  {journal} {Phys. Rev. Lett.}\ }\textbf
  {\bibinfo {volume} {126}},\ \bibinfo {pages} {076802} (\bibinfo {year}
  {2021})}\BibitemShut {NoStop}%
\bibitem [{\citenamefont {Khaymovich}\ \emph {et~al.}(2009)\citenamefont
  {Khaymovich}, \citenamefont {Kopnin}, \citenamefont {Mel'nikov},\ and\
  \citenamefont {Shereshevskii}}]{PhysRevB.79.224506}%
  \BibitemOpen
  \bibfield  {author} {\bibinfo {author} {\bibfnamefont {I.~M.}\ \bibnamefont
  {Khaymovich}}, \bibinfo {author} {\bibfnamefont {N.~B.}\ \bibnamefont
  {Kopnin}}, \bibinfo {author} {\bibfnamefont {A.~S.}\ \bibnamefont
  {Mel'nikov}},\ and\ \bibinfo {author} {\bibfnamefont {I.~A.}\ \bibnamefont
  {Shereshevskii}},\ }\bibfield  {title} {\bibinfo {title} {Vortex core states
  in superconducting graphene},\ }\href
  {https://doi.org/10.1103/PhysRevB.79.224506} {\bibfield  {journal} {\bibinfo
  {journal} {Phys. Rev. B}\ }\textbf {\bibinfo {volume} {79}},\ \bibinfo
  {pages} {224506} (\bibinfo {year} {2009})}\BibitemShut {NoStop}%
\bibitem [{\citenamefont {Caroli}\ \emph
  {et~al.}(1964{\natexlab{a}})\citenamefont {Caroli}, \citenamefont {{De
  Gennes}},\ and\ \citenamefont {Matricon}}]{CAROLI1964307}%
  \BibitemOpen
  \bibfield  {author} {\bibinfo {author} {\bibfnamefont {C.}~\bibnamefont
  {Caroli}}, \bibinfo {author} {\bibfnamefont {P.}~\bibnamefont {{De
  Gennes}}},\ and\ \bibinfo {author} {\bibfnamefont {J.}~\bibnamefont
  {Matricon}},\ }\bibfield  {title} {\bibinfo {title} {Bound fermion states on
  a vortex line in a type ii superconductor},\ }\href
  {https://doi.org/https://doi.org/10.1016/0031-9163(64)90375-0} {\bibfield
  {journal} {\bibinfo  {journal} {Physics Letters}\ }\textbf {\bibinfo {volume}
  {9}},\ \bibinfo {pages} {307} (\bibinfo {year}
  {1964}{\natexlab{a}})}\BibitemShut {NoStop}%
\bibitem [{\citenamefont {Chen}\ \emph {et~al.}(2020)\citenamefont {Chen},
  \citenamefont {Liu}, \citenamefont {Bao}, \citenamefont {Yan}, \citenamefont
  {Wang}, \citenamefont {Zhang},\ and\ \citenamefont
  {Feng}}]{PhysRevLett.124.097001}%
  \BibitemOpen
  \bibfield  {author} {\bibinfo {author} {\bibfnamefont {C.}~\bibnamefont
  {Chen}}, \bibinfo {author} {\bibfnamefont {Q.}~\bibnamefont {Liu}}, \bibinfo
  {author} {\bibfnamefont {W.-C.}\ \bibnamefont {Bao}}, \bibinfo {author}
  {\bibfnamefont {Y.}~\bibnamefont {Yan}}, \bibinfo {author} {\bibfnamefont
  {Q.-H.}\ \bibnamefont {Wang}}, \bibinfo {author} {\bibfnamefont
  {T.}~\bibnamefont {Zhang}},\ and\ \bibinfo {author} {\bibfnamefont
  {D.}~\bibnamefont {Feng}},\ }\bibfield  {title} {\bibinfo {title}
  {Observation of discrete conventional caroli--de gennes--matricon states in
  the vortex core of single-layer $\mathrm{FeSe}/{\mathrm{srtio}}_{3}$},\
  }\href {https://doi.org/10.1103/PhysRevLett.124.097001} {\bibfield  {journal}
  {\bibinfo  {journal} {Phys. Rev. Lett.}\ }\textbf {\bibinfo {volume} {124}},\
  \bibinfo {pages} {097001} (\bibinfo {year} {2020})}\BibitemShut {NoStop}%
\bibitem [{sup({\natexlab{a}})}]{supplemental}%
  \BibitemOpen
  \href@noop {} {} ({\natexlab{a}}),\ \bibinfo {note} {see Supplemental
  Material for details}\BibitemShut {NoStop}%
\bibitem [{\citenamefont {Liu}\ \emph {et~al.}(2016)\citenamefont {Liu},
  \citenamefont {Wang},\ and\ \citenamefont {Wang}}]{PhysRevB.93.035102}%
  \BibitemOpen
  \bibfield  {author} {\bibinfo {author} {\bibfnamefont {J.-G.}\ \bibnamefont
  {Liu}}, \bibinfo {author} {\bibfnamefont {D.}~\bibnamefont {Wang}},\ and\
  \bibinfo {author} {\bibfnamefont {Q.-H.}\ \bibnamefont {Wang}},\ }\bibfield
  {title} {\bibinfo {title} {Quantum impurities in channel mixing baths},\
  }\href {https://doi.org/10.1103/PhysRevB.93.035102} {\bibfield  {journal}
  {\bibinfo  {journal} {Phys. Rev. B}\ }\textbf {\bibinfo {volume} {93}},\
  \bibinfo {pages} {035102} (\bibinfo {year} {2016})}\BibitemShut {NoStop}%
\bibitem [{\citenamefont {Shirakawa}\ and\ \citenamefont
  {Yunoki}(2014)}]{PhysRevB.90.195109}%
  \BibitemOpen
  \bibfield  {author} {\bibinfo {author} {\bibfnamefont {T.}~\bibnamefont
  {Shirakawa}}\ and\ \bibinfo {author} {\bibfnamefont {S.}~\bibnamefont
  {Yunoki}},\ }\bibfield  {title} {\bibinfo {title} {Block lanczos
  density-matrix renormalization group method for general anderson impurity
  models: Application to magnetic impurity problems in graphene},\ }\href
  {https://doi.org/10.1103/PhysRevB.90.195109} {\bibfield  {journal} {\bibinfo
  {journal} {Phys. Rev. B}\ }\textbf {\bibinfo {volume} {90}},\ \bibinfo
  {pages} {195109} (\bibinfo {year} {2014})}\BibitemShut {NoStop}%
\bibitem [{\citenamefont {Allerdt}\ and\ \citenamefont
  {Feiguin}(2019)}]{allerdt_numerically_2019}%
  \BibitemOpen
  \bibfield  {author} {\bibinfo {author} {\bibfnamefont {A.}~\bibnamefont
  {Allerdt}}\ and\ \bibinfo {author} {\bibfnamefont {A.~E.}\ \bibnamefont
  {Feiguin}},\ }\bibfield  {title} {\bibinfo {title} {A numerically exact
  approach to quantum impurity problems in realistic lattice geometries},\
  }\bibfield  {journal} {\bibinfo  {journal} {Frontiers in Physics}\ }\textbf
  {\bibinfo {volume} {7}},\ \href {https://doi.org/10.3389/fphy.2019.00067}
  {10.3389/fphy.2019.00067} (\bibinfo {year} {2019}),\ \bibinfo {note}
  {publisher: Frontiers}\BibitemShut {NoStop}%
\bibitem [{\citenamefont {Yoshioka}\ and\ \citenamefont
  {Ohashi}(2000)}]{yoshioka2000numerical}%
  \BibitemOpen
  \bibfield  {author} {\bibinfo {author} {\bibfnamefont {T.}~\bibnamefont
  {Yoshioka}}\ and\ \bibinfo {author} {\bibfnamefont {Y.}~\bibnamefont
  {Ohashi}},\ }\bibfield  {title} {\bibinfo {title} {Numerical renormalization
  group studies on single impurity anderson model in superconductivity: A
  unified treatment of magnetic, nonmagnetic impurities, and resonance
  scattering},\ }\href@noop {} {\bibfield  {journal} {\bibinfo  {journal}
  {Journal of the Physical Society of Japan}\ }\textbf {\bibinfo {volume}
  {69}},\ \bibinfo {pages} {1812} (\bibinfo {year} {2000})}\BibitemShut
  {NoStop}%
\bibitem [{\citenamefont {Balatsky}\ \emph {et~al.}(2006)\citenamefont
  {Balatsky}, \citenamefont {Vekhter},\ and\ \citenamefont
  {Zhu}}]{RevModPhys.78.373}%
  \BibitemOpen
  \bibfield  {author} {\bibinfo {author} {\bibfnamefont {A.~V.}\ \bibnamefont
  {Balatsky}}, \bibinfo {author} {\bibfnamefont {I.}~\bibnamefont {Vekhter}},\
  and\ \bibinfo {author} {\bibfnamefont {J.-X.}\ \bibnamefont {Zhu}},\
  }\bibfield  {title} {\bibinfo {title} {Impurity-induced states in
  conventional and unconventional superconductors},\ }\href
  {https://doi.org/10.1103/RevModPhys.78.373} {\bibfield  {journal} {\bibinfo
  {journal} {Rev. Mod. Phys.}\ }\textbf {\bibinfo {volume} {78}},\ \bibinfo
  {pages} {373} (\bibinfo {year} {2006})}\BibitemShut {NoStop}%
\bibitem [{sup({\natexlab{b}})}]{supplemental2}%
  \BibitemOpen
  \href@noop {} {} ({\natexlab{b}}),\ \bibinfo {note} {it is also noted that
  the ground state here can adiabatically connect to that in the vortex free
  case without level crossing, see Supplementary Material for
  details}\BibitemShut {NoStop}%
\bibitem [{\citenamefont {Walsh}\ \emph {et~al.}(2019)\citenamefont {Walsh},
  \citenamefont {S\'emon}, \citenamefont {Poulin}, \citenamefont {Sordi},\ and\
  \citenamefont {Tremblay}}]{PhysRevLett.122.067203}%
  \BibitemOpen
  \bibfield  {author} {\bibinfo {author} {\bibfnamefont {C.}~\bibnamefont
  {Walsh}}, \bibinfo {author} {\bibfnamefont {P.}~\bibnamefont {S\'emon}},
  \bibinfo {author} {\bibfnamefont {D.}~\bibnamefont {Poulin}}, \bibinfo
  {author} {\bibfnamefont {G.}~\bibnamefont {Sordi}},\ and\ \bibinfo {author}
  {\bibfnamefont {A.-M.~S.}\ \bibnamefont {Tremblay}},\ }\bibfield  {title}
  {\bibinfo {title} {Local entanglement entropy and mutual information across
  the mott transition in the two-dimensional hubbard model},\ }\href
  {https://doi.org/10.1103/PhysRevLett.122.067203} {\bibfield  {journal}
  {\bibinfo  {journal} {Phys. Rev. Lett.}\ }\textbf {\bibinfo {volume} {122}},\
  \bibinfo {pages} {067203} (\bibinfo {year} {2019})}\BibitemShut {NoStop}%
\bibitem [{\citenamefont {Salomaa}(1988)}]{PhysRevB.37.9312}%
  \BibitemOpen
  \bibfield  {author} {\bibinfo {author} {\bibfnamefont {M.~M.}\ \bibnamefont
  {Salomaa}},\ }\bibfield  {title} {\bibinfo {title} {Schrieffer-wolff
  transformation for the anderson hamiltonian in a superconductor},\ }\href
  {https://doi.org/10.1103/PhysRevB.37.9312} {\bibfield  {journal} {\bibinfo
  {journal} {Phys. Rev. B}\ }\textbf {\bibinfo {volume} {37}},\ \bibinfo
  {pages} {9312} (\bibinfo {year} {1988})}\BibitemShut {NoStop}%
\bibitem [{\citenamefont {Zhang}\ \emph {et~al.}(2010)\citenamefont {Zhang},
  \citenamefont {Xie},\ and\ \citenamefont {Sun}}]{PhysRevB.82.075111}%
  \BibitemOpen
  \bibfield  {author} {\bibinfo {author} {\bibfnamefont {H.}~\bibnamefont
  {Zhang}}, \bibinfo {author} {\bibfnamefont {X.~C.}\ \bibnamefont {Xie}},\
  and\ \bibinfo {author} {\bibfnamefont {Q.-f.}\ \bibnamefont {Sun}},\
  }\bibfield  {title} {\bibinfo {title} {Scaling feature of magnetic field
  induced kondo-peak splittings},\ }\href
  {https://doi.org/10.1103/PhysRevB.82.075111} {\bibfield  {journal} {\bibinfo
  {journal} {Phys. Rev. B}\ }\textbf {\bibinfo {volume} {82}},\ \bibinfo
  {pages} {075111} (\bibinfo {year} {2010})}\BibitemShut {NoStop}%
\bibitem [{\citenamefont {Luo}\ \emph {et~al.}(2004)\citenamefont {Luo},
  \citenamefont {Xiang}, \citenamefont {Wang}, \citenamefont {Su},\ and\
  \citenamefont {Yu}}]{PhysRevLett.92.256602}%
  \BibitemOpen
  \bibfield  {author} {\bibinfo {author} {\bibfnamefont {H.~G.}\ \bibnamefont
  {Luo}}, \bibinfo {author} {\bibfnamefont {T.}~\bibnamefont {Xiang}}, \bibinfo
  {author} {\bibfnamefont {X.~Q.}\ \bibnamefont {Wang}}, \bibinfo {author}
  {\bibfnamefont {Z.~B.}\ \bibnamefont {Su}},\ and\ \bibinfo {author}
  {\bibfnamefont {L.}~\bibnamefont {Yu}},\ }\bibfield  {title} {\bibinfo
  {title} {Fano resonance for anderson impurity systems},\ }\href
  {https://doi.org/10.1103/PhysRevLett.92.256602} {\bibfield  {journal}
  {\bibinfo  {journal} {Phys. Rev. Lett.}\ }\textbf {\bibinfo {volume} {92}},\
  \bibinfo {pages} {256602} (\bibinfo {year} {2004})}\BibitemShut {NoStop}%
\bibitem [{\citenamefont {Vernek}\ \emph {et~al.}(2014)\citenamefont {Vernek},
  \citenamefont {Penteado}, \citenamefont {Seridonio},\ and\ \citenamefont
  {Egues}}]{PhysRevB.89.165314}%
  \BibitemOpen
  \bibfield  {author} {\bibinfo {author} {\bibfnamefont {E.}~\bibnamefont
  {Vernek}}, \bibinfo {author} {\bibfnamefont {P.~H.}\ \bibnamefont
  {Penteado}}, \bibinfo {author} {\bibfnamefont {A.~C.}\ \bibnamefont
  {Seridonio}},\ and\ \bibinfo {author} {\bibfnamefont {J.~C.}\ \bibnamefont
  {Egues}},\ }\bibfield  {title} {\bibinfo {title} {Subtle leakage of a
  majorana mode into a quantum dot},\ }\href
  {https://doi.org/10.1103/PhysRevB.89.165314} {\bibfield  {journal} {\bibinfo
  {journal} {Phys. Rev. B}\ }\textbf {\bibinfo {volume} {89}},\ \bibinfo
  {pages} {165314} (\bibinfo {year} {2014})}\BibitemShut {NoStop}%
\bibitem [{\citenamefont {Ruiz-Tijerina}\ \emph {et~al.}(2015)\citenamefont
  {Ruiz-Tijerina}, \citenamefont {Vernek}, \citenamefont {Dias~da Silva},\ and\
  \citenamefont {Egues}}]{PhysRevB.91.115435}%
  \BibitemOpen
  \bibfield  {author} {\bibinfo {author} {\bibfnamefont {D.~A.}\ \bibnamefont
  {Ruiz-Tijerina}}, \bibinfo {author} {\bibfnamefont {E.}~\bibnamefont
  {Vernek}}, \bibinfo {author} {\bibfnamefont {L.~G. G.~V.}\ \bibnamefont
  {Dias~da Silva}},\ and\ \bibinfo {author} {\bibfnamefont {J.~C.}\
  \bibnamefont {Egues}},\ }\bibfield  {title} {\bibinfo {title} {Interaction
  effects on a majorana zero mode leaking into a quantum dot},\ }\href
  {https://doi.org/10.1103/PhysRevB.91.115435} {\bibfield  {journal} {\bibinfo
  {journal} {Phys. Rev. B}\ }\textbf {\bibinfo {volume} {91}},\ \bibinfo
  {pages} {115435} (\bibinfo {year} {2015})}\BibitemShut {NoStop}%
\bibitem [{\citenamefont {Hosur}\ \emph {et~al.}(2011)\citenamefont {Hosur},
  \citenamefont {Ghaemi}, \citenamefont {Mong},\ and\ \citenamefont
  {Vishwanath}}]{PhysRevLett.107.097001}%
  \BibitemOpen
  \bibfield  {author} {\bibinfo {author} {\bibfnamefont {P.}~\bibnamefont
  {Hosur}}, \bibinfo {author} {\bibfnamefont {P.}~\bibnamefont {Ghaemi}},
  \bibinfo {author} {\bibfnamefont {R.~S.~K.}\ \bibnamefont {Mong}},\ and\
  \bibinfo {author} {\bibfnamefont {A.}~\bibnamefont {Vishwanath}},\ }\bibfield
   {title} {\bibinfo {title} {Majorana modes at the ends of superconductor
  vortices in doped topological insulators},\ }\href
  {https://doi.org/10.1103/PhysRevLett.107.097001} {\bibfield  {journal}
  {\bibinfo  {journal} {Phys. Rev. Lett.}\ }\textbf {\bibinfo {volume} {107}},\
  \bibinfo {pages} {097001} (\bibinfo {year} {2011})}\BibitemShut {NoStop}%
\bibitem [{\citenamefont {Caroli}\ \emph
  {et~al.}(1964{\natexlab{b}})\citenamefont {Caroli}, \citenamefont
  {De~Gennes},\ and\ \citenamefont {Matricon}}]{caroli_bound_1964}%
  \BibitemOpen
  \bibfield  {author} {\bibinfo {author} {\bibfnamefont {C.}~\bibnamefont
  {Caroli}}, \bibinfo {author} {\bibfnamefont {P.~G.}\ \bibnamefont
  {De~Gennes}},\ and\ \bibinfo {author} {\bibfnamefont {J.}~\bibnamefont
  {Matricon}},\ }\bibfield  {title} {\bibinfo {title} {Bound {Fermion} states
  on a vortex line in a type ii superconductor},\ }\href
  {https://doi.org/10.1016/0031-9163(64)90375-0} {\bibfield  {journal}
  {\bibinfo  {journal} {Physics Letters}\ }\textbf {\bibinfo {volume} {9}},\
  \bibinfo {pages} {307} (\bibinfo {year} {1964}{\natexlab{b}})}\BibitemShut
  {NoStop}%
\end{thebibliography}%

\end{document}